\documentclass[a4paper,11pt]{article}
\usepackage{latexsym}
\usepackage{amssymb}
\usepackage{amsmath}
\usepackage{cite}
\usepackage{color}

\usepackage{microtype}

\numberwithin{equation}{section}
\allowdisplaybreaks

\makeatletter \@addtoreset{equation}{section} \makeatother

\usepackage{ytableau}

\ytableausetup{centertableaux, mathmode, smalltableaux}

\definecolor{blue-violet}{rgb}{0.54, 0.17, 0.89}
\definecolor{PineGreen}{cmyk}{0.92, 0, 0.59, 0.25}
\definecolor{YellowOrange}{cmyk}{0, 0.42, 1, 0}

\newcommand{\be}{\begin{equation}}
\newcommand{\ee}{\end{equation}}
\newcommand{\beq} {\begin{equation}}
\newcommand{\eeq} {\end{equation}}
\newcommand{\ba}{\begin{eqnarray}}
\newcommand{\ea}{\end{eqnarray}}

\usepackage{graphicx}

\begin{document}
\numberwithin{equation}{section}

\begin{center}
{\bf\LARGE Metric-Affine Vector-Tensor Correspondence and Implications in $F(R,T,Q,\mathcal{T},\mathcal{D})$ gravity} \\
\vskip 2 cm
{\bf \large Damianos Iosifidis$^{1}$, Ratbay Myrzakulov$^{2,3}$, Lucrezia Ravera$^{4,5}$, Gulmira Yergaliyeva$^{2,3}$, Koblandy Yerzhanov$^{2,3}$}
\vskip 8mm
 \end{center}
\noindent {\small $^{1}$ \it Institute of Theoretical Physics, Department of Physics, Aristotle University of Thessaloniki, 54124 Thessaloniki, Greece. \\
$^{2}$ \it Ratbay Myrzakulov Eurasian International Centre for Theoretical Physics, Nur-Sultan, 010009, Kazakhstan. \\
$^{3}$ \it Eurasian National University, Nur-Sultan, 010008, Kazakhstan. \\
$^{4}$ \it DISAT, Politecnico di Torino, Corso Duca degli Abruzzi 24, 10129 Torino, Italy. \\
$^{5}$  \it INFN, Sezione di Torino, Via P. Giuria 1, 10125 Torino, Italy.
}

\begin{center}
\today
\end{center}

\vskip 2 cm
\begin{center}
{\small {\bf Abstract}}
\end{center}

We extend the results of antecedent literature on quadratic Metric-Affine Gravity by studying a new quadratic gravity action in vacuum which, besides the usual (non-Riemannian) Einstein-Hilbert contribution, involves all the parity even quadratic terms in torsion and non-metricity plus a Lagrangian that is quadratic in the field-strengths of the torsion and non-metricity vector traces. The theory result to be equivalent, on-shell, to a Vector-Tensor theory. We also discuss the sub-cases in which the contribution to the Lagrangian quadratic in the field-strengths of the torsion and non-metricity vectors just exhibits one of the aforementioned quadratic terms. We then report on implications of our findings in the context of $F(R,T,Q,\mathcal{T},\mathcal{D})$ gravity.

\vfill
\noindent {\small{\it
    E-mail:  \\
{\tt diosifid@auth.gr}; \\
{\tt rmyrzakulov@gmail.com}; \\
{\tt lucrezia.ravera@polito.it}; \\
{\tt gyergaliyeva@gmail.com}; \\
{\tt yerzhanovkk@gmail.com}.}}
   \eject

\tableofcontents

\noindent\hrulefill

\section{Introduction}\label{intro}

The emergence of non-Euclidean geometry led to extraordinary progresses in Mathematics and Physics. In particular, the development of Riemannian geometry allowed the rigorous mathematical formulation of general relativity (GR), Einstein's well-celebrated theory of gravity. Nevertheless, the latter is not devoid of limitations and open questions. In these regards, diverse alternative/modified theories of gravity have been proposed. A propitious setup in the spirit of gravity geometrization is that of non-Riemannian geometry \cite{eisenhart,schouten}, where the Riemannian assumptions of metric compatibility and torsionlessness of the connection are released.
The inclusion of torsion and non-metricity in gravity models has provided useful applications in various mathematical and physical contexts, ranging from a clearer understanding of the geometrical features of manifolds involving torsion and non-metricity to cosmological aspects of alternative/modified theories of gravity, among which, e.g., the results presented in \cite{Klemm:2018bil,Kranas:2018jdc,Iosifidis:2018zjj,Iosifidis:2018zwo,Iosifidis:2018jwu,Klemm:2019izb,Pereira:2019yhu,Barrow:2019bvx,Jimenez:2019woj,BeltranJimenez:2019odq,Iosifidis:2019fsh,Klemm:2020mfp,Klemm:2020gfm,Iosifidis:2020dck,Guimaraes:2020drj,Iosifidis:2020gth,Iosifidis:2020zzp,Iosifidis:2021iuw,Saridakis:2021lqd,Iosifidis:2021kqo,Myrzakulov:2021vel,Harko:2014sja,Harko:2014aja,Harko:2021bdi,Saridakis:2019qwt}. The literature on the subject is huge, and here we report just some recent developments.
Moreover, non-Riemannian geometry constitues the geometric setup underlying Metric-Affine Gravity (MAG) \cite{Hehl:1994ue,Gronwald:1997bx,Hehl:1999sb,Iosifidis:2019jgi,Percacci:2020ddy,Shimada:2018lnm,Delhom:2019zrb,Hohmann:2019fvf,Mikura:2020qhc,BeltranJimenez:2020sqf}, a particularly promising and rather general framework in which the metric and the general affine connection are considered, a priori, as independent fields and the matter Lagrangian depends on the connection as well, which leads to introduce the so-called hypermomentum tensor \cite{Hehl:1976kt,Hehl:1976kt2,Hehl:1976kv}. The extensive study of MAG has started a few decades ago and it is now known that many alternative theories of gravity can be obtained as special cases of MAG. Nevertheless, there are some open issues regarding the very general MAG setup that still need and deserve to be addressed, whose understanding could provide remarkable insights in the context of alternative theories of gravity, where torsion and non-metricity may also play a key role in explaining cosmological and astrophysical aspects of our Universe, such as the origin of dark matter and dark energy. 

With this in mind, in the present work we extend some results of antecedent literature \cite{Tucker:1995fw,Obukhov:1996pf,Vlachynsky:1996zh,Dereli:1996ex,Obukhov:1996ka,Obukhov:1997zd,Vitagliano:2010sr,Baekler:2011jt,Vitagliano:2013rna,Iosifidis:2021fnq} by studying a general quadratic MAG action in vacuum (namely in the absence of matter) in $n$ spacetime dimensions. More precisely, in \cite{Obukhov:1996ka,Obukhov:1997zd} the authors examined a quadratic theory involving also a piece quadratic in the so-called homothetic curvature tensor, while in \cite{Vitagliano:2010sr} only contributions quadratic in the torsion tensor were considered. 
The theory we consider here involves, besides the usual (non-Riemannian) Einstein-Hilbert contribution, all the parity even quadratic terms in torsion and non-metricity plus a Lagrangian quadratic in the field-strengths of the torsion and non-metricity vectors. The gravitational action represents a generalization of the one studied in \cite{Obukhov:1996ka,Obukhov:1997zd} and an extension of the quadratic model analyzed in \cite{Iosifidis:2021fnq} by including quadratic contributions in all the aforementioned field-strengths.
The paper is organized as follows: In Section \ref{notconv} we review the geometric framework, adopting the same notation and conventions of \cite{Iosifidis:2019jgi}. In Section \ref{thetheory} we write the action of the model and derive its field equations. Consequently, we consider the sub-cases in which the contribution to the Lagrangian quadratic in the field-strengths of the torsion and non-metricity vectors just exhibits one of the aforementioned quadratic terms. By analyzing the full quadratic theory we show that the latter propagates three additional degrees of freedom in comparison to GR. In particular we show that the quadratic theory (with the kinetic terms included) is on-shell equaivallent to GR with three interacting Proca fields propagating in spacetime. The masses of these Proca fields and the strengths of their interactions depend on the parameters of the theory. Continuing, in Section \ref{Myrzgrav} we provide implications of the formulation to the case of (linear) $F(R,T,Q,\mathcal{T},\mathcal{D})$ gravity, also known in the literature as Metric-Affine Myrzakulov Gravity VIII (MAMG-VIII), in vacuum, see, e.g., \cite{Iosifidis:2021kqo} and the recent review \cite{Myrzakulov:2021vel}. Section \ref{concl} is devoted to some final remarks and possible future developments. In Appendix \ref{appa} we collect some useful formulas and results of intermediate calculations.

\section{Torsion and non-metricity decomposition}\label{notconv}

We consider $n$ spacetime dimensions. Our metric convention is mostly plus. A general affine connection can be decomposed as ($\mu, \nu,\ldots=0,1,\ldots,n$)
\begin{equation}\label{gendecompaffconn}
{\Gamma^\lambda}_{\mu\nu} = \tilde{\Gamma}^\lambda_{\phantom{\lambda}\mu\nu} + {N^\lambda}_{\mu\nu}\,,
\end{equation}
where
\begin{equation}\label{distortion}
{N^\lambda}_{\mu\nu} = \underbrace{\frac12 g^{\rho\lambda}\left(Q_{\mu\nu\rho} + Q_{\nu\rho\mu}
- Q_{\rho\mu\nu}\right)}_{\text{deflection {(or disformation)}}} - \underbrace{g^{\rho\lambda}\left(S_{\rho\mu\nu} +
S_{\rho\nu\mu} - S_{\mu\nu\rho}\right)}_{\text{contorsion} \, := \, {K^\lambda}_{\mu \nu}}
\end{equation}
is the distortion tensor and 
\begin{equation}\label{lcconn}
\tilde{\Gamma}^\lambda_{\phantom{\lambda}\mu\nu} = \frac12 g^{\rho\lambda}\left(\partial_\mu 
g_{\nu\rho} + \partial_\nu g_{\rho\mu} - \partial_\rho g_{\mu\nu}\right)
\end{equation}
is the Levi-Civita connection. The torsion and non-metricity tensors in \eqref{distortion} are respectively defined as follows:
\begin{equation}
\begin{split}
{S_{\mu\nu}}^\rho & := {\Gamma^\rho}_{[\mu\nu]}\,, \\ 
Q_{\lambda\mu\nu} & := -\nabla_\lambda g_{\mu\nu} = 
-\partial_\lambda g_{\mu\nu} + {\Gamma^\rho}_{\mu\lambda} g_{\rho\nu} +
{\Gamma^\rho}_{\nu\lambda}g_{\mu\rho} \,,
\end{split}
\end{equation}
where $\nabla$ denotes the covariant derivative associated with the general affine connection $\Gamma$.
Here, let us recall the trace decomposition of torsion and non-metricity, since it will be particularly useful in the following. In $n$ dimensions it reads
\begin{equation}\label{dectorandnm}
\begin{split}
{S_{\lambda\mu}}^\nu & = \frac{2}{1-n} {\delta_{[\lambda}}^{\nu} S_{\mu]} + {T_{\lambda\mu}}^\nu \,,  \\
Q_{\alpha\mu\nu} & = \frac{\left[(n+1)Q_\alpha - 2 q_\alpha \right]}{(n+2)(n-1)} g_{\mu \nu} + \frac{2 \left[n q_{(\mu}g_{\nu)\alpha} - Q_{(\mu}g_{\nu)\alpha}\right]}{(n+2)(n-1)} + \Omega_{\lambda\mu\nu}\,, 
\end{split}
\end{equation}
where $Q_\lambda := {Q_{\lambda \mu}}^\mu$ and $q_\nu := {Q^\mu}_{\mu\nu}$ are the non-metricity vectors and $S_\lambda :={S_{\lambda \sigma}}^{\sigma}$ is the torsion vector, while ${T_{\lambda\mu}}^\nu$ and $\Omega_{\lambda\mu\nu}$ are the traceless parts of torsion and non-metricity, respectively.
We adopt the following definition of the Riemann tensor for a general affine connection ${\Gamma^\lambda}_{\mu \nu}$:
\begin{equation}\label{defRiem}
{R^\mu}_{\nu \alpha \beta} := 2 \partial_{[\alpha} {\Gamma^\mu}_{|\nu|\beta]} + 2 {\Gamma^\mu}_{\rho [\alpha} {\Gamma^\rho}_{|\nu |\beta]} \,.
\end{equation}
Correspondingly, $R_{\mu \nu}={R^\rho}_{\mu \rho \nu}$ and $R=g^{\mu \nu} R_{\mu \nu}$ are, respectively, the Ricci tensor and the scalar curvature of $\Gamma$.
Let us also mention that in the following we will need the variation of the torsion and non-metricity with respect to the metric tensor and the general affine connection ${\Gamma^\lambda}_{\mu \nu}$, namely \cite{Iosifidis:2019jgi}
\begin{equation}\label{variationQS}
\begin{split}
& \delta_g Q_{\rho\alpha\beta} = \partial_\rho\left(g_{\mu\alpha} g_{\nu\beta} \delta g^{\mu\nu}\right) - 
2 g_{\lambda\mu} g_{\nu(\alpha} {\Gamma^\lambda}_{\beta)\rho}\delta g^{\mu\nu}\,, \\
& \delta_g {S_{\mu\nu}}^\alpha = 0 \,, \\
& \delta_\Gamma Q_{\rho\alpha\beta} = 2\delta^\nu_\rho\delta^\mu_{(\alpha} g_{\beta)\lambda}
\delta {\Gamma^\lambda}_{\mu\nu}\,, \\
& \delta_\Gamma {S_{\alpha\beta}}^\lambda = \delta^{[\mu}_{\alpha}\delta^{\nu]}_{\beta}
\delta {\Gamma^\lambda}_{\mu\nu}\,.
\end{split}
\end{equation}
Then, from \eqref{variationQS}, one can derive the variation of the torsion and non-metricity vectors with respect to the metric and the general affine connection (see \cite{Iosifidis:2019jgi}).

\section{The quadratic theory}\label{thetheory}

We shall start by considering the following action in $n$ dimensions:
\begin{equation}\label{thy}
\begin{split}
S[g,\Gamma] &=\frac{1}{2 \kappa}\int d^{n}x \sqrt{-g} \Big[ R+  
b_{1}S_{\alpha\mu\nu}S^{\alpha\mu\nu} +
b_{2}S_{\alpha\mu\nu}S^{\mu\nu\alpha} +
b_{3}S_{\mu}S^{\mu} \\
& + a_{1}Q_{\alpha\mu\nu}Q^{\alpha\mu\nu} +
a_{2}Q_{\alpha\mu\nu}Q^{\mu\nu\alpha} +
a_{3}Q_{\mu}Q^{\mu}+
a_{4}q_{\mu}q^{\mu}+
a_{5}Q_{\mu}q^{\mu} \\
& +c_{1}Q_{\alpha\mu\nu}S^{\alpha\mu\nu}+
c_{2}Q_{\mu}S^{\mu} +
c_{3}q_{\mu}S^{\mu}+\mathcal{L}_{\text{FS}} \Big] \\
& =\frac{1}{2 \kappa}\int d^{n}x \sqrt{-g} \Big[ R+ \mathcal{L}_{2}+\mathcal{L}_{\text{FS}} \Big] \,,
\end{split}
\end{equation}
where $R$ is the scalar curvature of the general affine connection ${\Gamma^\lambda}_{\mu \nu}$, $\mathcal{L}_2$ contains all the torsion and non-metricity scalar terms and
\beq\label{LFS}
\mathcal{L}_{\text{FS}}=k_{1}\hat{R}_{\mu\nu}\hat{R}^{\mu\nu}+k_{2}q_{\mu\nu}q^{\mu\nu}+k_{3}S_{\mu\nu}S^{\mu\nu}
\eeq
is the Lagrangian contribution given in terms of the field-strengths
\beq\label{fieldstr}
\hat{R}_{\mu \nu}:= \partial_{[\mu} Q_{\nu]}\,, \quad q_{\mu\nu}:=\partial_{[\mu}q_{\nu]} \,, \quad S_{\mu\nu}:=\partial_{[\mu}S_{\nu]} \,.
\eeq
Here, $\hat{R}_{\mu \nu}$ the homothetic curvature tensor, while $q_{\mu\nu}$ and $S_{\mu \nu}$ are the field-strengths of $q_\mu$ and $S_\mu$, respectively.

Some comments are in order regarding $\mathcal{L}_2$. In fact, notice that, restricting to $\mathcal{L}_2$, for the parameter choice $b_1 = 1$, $b_2 = -2$, $b_3 =- 4$, $a_i= c_j= 0$ ($i=1,2,\ldots,5$, $j=1,2,3$) and
imposing vanishing curvature and zero non-metricity, one recovers the teleparallel equivalent
of GR. In addition, demanding vanishing curvature and zero torsion and taking $a_1 = -a_3 = \frac{1}{4}$, $a_2 = -a_5 = -\frac{1}{2}$, $a_4 = 0$, $b_j =c_j= 0$, $\mathcal{L}_2$ reduces to the symmetric teleparallel equivalent of
GR. Furthermore, if we pick $b_1 = 1$, $b_2 = -1$, $b_3 = -4$, $a_1 = -a_3 = \frac{1}{4}$, $a_2 = -a_5 = - \frac{1}{2}$, $a_4=0$, $c_1 = -c_2 = c_3 = 2$ and impose only vanishing curvature, $\mathcal{L}_2$ boils down to a generalized equivalent to GR that admits both torsion and non-metricity.
For the sake of generality, however, here we will not make any assumptions about the parameters of the theory and will consider the complete action \eqref{thy}.
We now move on to the study of the field equations of the theory by varying \eqref{thy} with respect to the general affine connection ${\Gamma^\lambda}_{\mu \nu}$ and the metric tensor, independently.

The connection field equations of \eqref{thy} read
\beq
{P_{\lambda}}^{\mu\nu}+{\Psi_{\lambda}}^{\mu\nu}+{B_{\lambda}}^{\mu\nu}=0 \,, \label{feGamma}
\eeq
being ${P_{\lambda}}^{\mu\nu}$ the Palatini tensor, which can be written in terms of torsion and non-metricity \cite{Iosifidis:2019jgi}, and where we have defined
\begin{gather}
{\Psi_{\lambda}}^{\mu\nu}:= {H^{\mu\nu}}_{\lambda}+\delta^{\mu}_{\lambda}k^{\nu}+\delta^{\nu}_{\lambda}h^{\mu}+g^{\mu\nu}h_{\lambda}+f^{[\mu}\delta^{\nu ]}_{\lambda} \,,
\end{gather}
with
\begin{equation}
{H^{\mu\nu}}_{\lambda} := 4 a_{1}{Q^{\nu\mu}}_{\lambda}+2 a_{2}({Q^{\mu\nu}}_{\lambda}+{Q_{\lambda}}^{\mu\nu})+2 b_{1}{S^{\mu\nu}}_{\lambda} + 2 b_{2}{S_{\lambda}}^{[\mu\nu]}+c_{1}( {S^{\nu\mu}}_{\lambda}-{S_{\lambda}}^{\nu\mu}+{Q^{[\mu\nu]}}_{\lambda}) \,,
\end{equation}
\begin{equation}
\begin{split}
k_{\mu} & = 4 a_{3}Q_{\mu}+2 a_{5}q_{\mu}+2 c_{2}S_{\mu} \,, \\
h_{\mu} &= a_{5} Q_{\mu}+2 a_{4}q_{\mu}+c_{3}S_{\mu} \,, \\
f_{\mu} & = c_{2} Q_{\mu}+ c_{3}q_{\mu}+2 b_{3}S_{\mu} \,,
\end{split}
\end{equation}
together with
\begin{equation}\label{Bmn}
{B_{\lambda}}^{\mu\nu}:=\frac{1}{\sqrt{-g}}\frac{\delta \left( \sqrt{-g} \mathcal{L}_{\text{FS}} \right)}{\delta {\Gamma^{\lambda}}_{\mu\nu}}= -2\Big[ 2 k_{1}\delta_{\lambda}^{\mu}(D_{\alpha}\hat{R}^{\alpha\nu})+k_{2}(g^{\mu\nu}g_{\beta\lambda}+\delta_{\beta}^{\mu}\delta_{\lambda}^{\nu})(D_{\alpha}q^{\alpha\beta})+k_{3}(D_{\alpha}S^{\alpha[\mu}\delta^{\nu]}_{\lambda})\Big] \,,
\end{equation}
where we have also introduced the operator
\beq
D_{\alpha} (\ldots):=\frac{1}{\sqrt{-g}}\partial_{\alpha}(\sqrt{-g} (\ldots)) \,.
\eeq
Then, let us compute the different traces of \eqref{feGamma}. For the ${\Psi_\lambda}^{\mu \nu}$ sector we get
\begin{equation}
\begin{split}
\Psi_{(1)}^{\nu} :={\Psi_{\mu}}^{\mu\nu} & =\Big[ 4a_{1}-\frac{c_{1}}{2}+4 n a_{3}+2 a_{5}+\frac{(1-n)}{2}c_{2} \Big] Q^{\nu} \\
& +\Big[ 4 a_{2}+\frac{c_{1}}{2}+2 n a_{5}+4 a_{4}+\frac{(1-n)}{2}c_{3} \Big] q^{\nu} \\
& +\Big[ -2b_{1}+b_{2}+2 c_{1}+2 n c_{2}+2 c_{3}+(1-n)b_{3} \Big] S^{\nu} \,,
\end{split}
\end{equation}
\begin{equation}
\begin{split}
\Psi_{(2)}^{\mu} :={\Psi_{\nu}}^{\mu\nu} & =\Big[2 a_{2}+\frac{c_{1}}{2}+4 a_{3}+(n+1) a_{5}+\frac{(n-1)}{2}c_{2} \Big] Q^{\nu} \\
& +\Big[ 4 a_{1}+2 a_{2}-\frac{c_{1}}{2}+2 a_{5}+2(n+1) a_{4}+\frac{(n-1)}{2}c_{3} \Big] q^{\nu} \\
& +\Big[ 2b_{1}-b_{2}- c_{1}+2  c_{2}+(n+1) c_{3}+ (n-1) b_{3} \Big] S^{\nu} \,,
\end{split}
\end{equation}
\begin{equation}
\begin{split}
\Psi_{(3)}^{\lambda} :=g^{\lambda \rho}{\Psi_{\rho}}^{\mu\nu} g_{\mu \nu} & =\Big[2 a_{2}+4 a_{3}+(n+1)a_{5}\Big]Q^{\lambda} +2\Big[2 a_{1}+a_{2}+(n+1)a_{4}+a_{5}\Big]q^{\lambda} \\
& +\Big[2 c_{2}-c_{1} +(n+1)c_{3}\Big]S^{\lambda} \,.
\end{split}
\end{equation}
Analogously, for ${B_{\lambda}}^{\mu\nu}$ we find
\beq
B_{(1)}^{\nu}=-2 D_{\alpha}\Big( 2 k_{1}n \hat{R}^{\alpha \nu}+2 k_{2}q^{\alpha \nu}+\frac{(1-n)}{2} k_3 S^{\alpha \nu}\Big) \,,
\eeq
\beq
B_{(2)}^{\mu}=-2 D_{\alpha}\Big( 2 k_{1} \hat{R}^{\alpha\mu}+ (n+1) k_{2}q^{\alpha\mu}+\frac{(n-1)}{2}k_3 S^{\alpha\mu}\Big) \,,
\eeq
\beq
B_{(3)}^{\lambda}=-2 D_{\alpha}\Big( 2 k_{1} \hat{R}^{\alpha \lambda}+ (n+1) k_{2}q^{\alpha \lambda}\Big) \,.
\eeq
Concerning the trace contributions from the Palatini tensor we have (see, e.g., \cite{Iosifidis:2019jgi})
\begin{equation}
P_{(1)}^\nu = 0 \,,
\end{equation}
\begin{equation}
P_{(2)}^\mu = \frac{1}{2} (1-n) Q^\mu + (n-1) q^\mu + 2 (2-n) S^\mu \,,
\end{equation}
\begin{equation}
P_{(3)}^\lambda = \frac{1}{2} (n-3) Q^\lambda + q^\lambda + 2 (n-2) S^\lambda \,.
\end{equation}
Therefore, taking the traces of \eqref{feGamma} and using all the above, we obtain the following set of equations:
\begin{equation}\label{e1ap}
\begin{split}
& N_1 Q^{\nu} + N_2 q^{\nu} + N_3 S^{\nu} = 2 D_{\alpha}\Big( 2 k_{1}n \hat{R}^{\alpha \nu}+2 k_{2}q^{\alpha \nu }+\frac{(1-n)}{2} k_3 S^{\alpha \nu}\Big) \,, \\
& N_4 Q^{\mu} + N_5 q^{\mu} + N_6 S^{\mu} = 2 D_{\alpha}\Big( 2 k_{1} \hat{R}^{\alpha\mu}+ (n+1) k_{2}q^{\alpha\mu}+\frac{(n-1)}{2} k_3 S^{\alpha\mu}\Big) \,, \\
& N_7 Q^{\lambda} + N_8 q^{\lambda} + N_9 S^{\lambda} = 2 D_{\alpha}\Big( 2 k_{1} \hat{R}^{\alpha \lambda}+ (n+1) k_{2}q^{\alpha \lambda}\Big) \,,
\end{split}
\end{equation}
where the explicit expressions of the coefficients $N_1,N_2,\ldots, N_9$ in terms of the parameters of the theory are collected in Appendix \ref{appa}.
We may now express the above system in matrix form according to
\beq
A \vec{X}=B \vec{Y} \,, \label{A}
\eeq
where $\vec{X}=(Q^{\mu},q^\mu,S^{\mu})^{T}$ and $\vec{Y}=(D_{\alpha}\hat{R}^{\alpha\mu},D_{\alpha}q^{\alpha\mu},D_{\alpha}S^{\alpha\mu})^{T}$ and $A$ and $B$ are the coefficient matrices of the system of equations. Assuming the general setting in which $k_1 \neq 0$, $k_2 \neq 0$, $k_3 \neq 0$ (together with $n \neq 1$ and the obvious $n \neq -2$), the determinant of $B$ is non-vanishing and therefore the inverse $B^{-1}$ exists. Then, by formally multiplying \eqref{A} with $B^{-1}$ from the left, we get
\beq
\vec{Y}=C \vec{X} \,, 
\eeq
with $C=B^{-1}A$.
More explicitly, this means that we are left with the following Proca-like equations for the torsion and non-metricity vectors:
\begin{equation}\label{procalike}
\begin{split}
D_\alpha \hat{R}^{\alpha \mu} & = c_{11} Q^\mu + c_{12} q^\mu + c_{13} S^\mu \,, \\
D_\alpha q^{\alpha \mu} & = c_{21} Q^\mu + c_{22} q^\mu + c_{23} S^\mu \,, \\
D_\alpha S^{\alpha \mu} & = c_{31} Q^\mu + c_{32} q^\mu + c_{33} S^\mu \,,
\end{split}
\end{equation}
where the $c$ coefficients are given in terms of the parameters of the theory in Appendix \ref{appa}.
Such coefficients play the role of (squared) masses in the Proca-like equations \eqref{procalike}.
Consequently, using \eqref{dectorandnm} and plugging back this result into \eqref{feGamma}, the latter boils down to
\begin{equation}\label{feGamma1}
(4 a_2 -2) {\Omega_\lambda}^{\mu \nu} + (4 a_2 + c_1) \Omega^{\mu \phantom{\lambda} \nu}_{\phantom{\mu}\lambda} + ( 8 a_1 - c_1 ) \Omega^{\nu \phantom{\lambda} \mu}_{\phantom{\nu}\lambda} + 2 (b_2 -2) {T_\lambda}^{\mu \nu} + 2(2 b_1 - c_1) {T^{\mu \nu}}_\lambda - 2 (b_2 + c_1) {T_\lambda}^{\nu \mu} = 0 \,,
\end{equation}
which relates the traceless parts of non-metricity and torsion. 
Taking the completely antisymmetric part of \eqref{feGamma1} we get
\begin{equation}
(b_1+b_2-1) T_{[\lambda \mu \nu]} = 0 \quad \Rightarrow \quad b_1+b_2-1 = 0 \quad \vee \quad T_{[\lambda \mu \nu]} = 0 \,,
\end{equation}
while considering the completely symmetric part of \eqref{feGamma1} we obtain
\begin{equation}
(4a_1 + 4a_2 - 1) \Omega_{(\lambda \mu \nu)} = 0 \quad \Rightarrow \quad 4a_1 + 4a_2 - 1 = 0 \quad \vee \quad \Omega_{(\lambda \mu \nu)} = 0 \,.
\end{equation}
Notice, in particular, that \eqref{feGamma1} can be solved by requiring
\begin{equation}\label{vanishingOmT}
\Omega_{\lambda \mu \nu} = 0 \,, \quad {T_{\lambda \mu}}^\nu = 0 \,,
\end{equation}
without binding any parameters.
On the other hand, it can also be solved by fixing the parameters as follows:
\begin{equation}
a_1 = - \frac{1}{4} \,, \quad a_2 = \frac{1}{2} \,, \quad b_1 = -1 \,, \quad b_2 = 2 \,, \quad c_1 = -2 \,,
\end{equation}
without constraining $\Omega_{\lambda \mu \nu}$ and ${T_{\lambda \mu}}^\nu$.
However, the latter would just fix five out of the fourteen parameters of the theory, namely those that do not explicitly involve the torsion a non-metricity vectors in \eqref{thy}. 
But this would mean having some specific symmetry underlying the theory from the start, therefore we simply neglect this case. Moreover, the solution \eqref{vanishingOmT} appears more appealing, as it reduces the number of degrees of freedom of the theory, eliminating the traceless components of torsion and non-metricity, in such a way that we are just left with the Proca-like equations \eqref{procalike} involving $Q_\mu$, $q_{\mu}$, and $S_\mu$.

Let us now move on to the variation of \eqref{thy} with respect to $g^{\mu \nu}$. Before proceeding, we define, on the same lines of \cite{Iosifidis:2019jgi}, the following quantities (also referred to as ``superpotentials''):
\begin{equation}
\begin{split}
\Xi^{\alpha \mu \nu} & := a_1 Q^{\alpha \mu \nu} + a_2 Q^{\mu \nu \alpha} + a_3 g^{\mu \nu} Q^\alpha + a_4 g^{\alpha \mu} q^\nu + a_5 g^{\alpha \mu} Q^\nu \,, \\
\Sigma^{\alpha \mu \nu} & := b_1 S^{\alpha \mu \nu} + b_2 S^{\mu \nu \alpha} + b_3 g^{\mu \nu} S^\alpha \,, \\
\Pi^{\alpha \mu \nu} & := c_1 S^{\alpha \mu \nu} + c_2 g^{\mu \nu} S^\alpha + c_3 g^{\alpha \mu} S^\nu \,,
\end{split}
\end{equation}
together with
\begin{equation}
\begin{split}
\mathcal{L}_Q & := Q_{\alpha \mu \nu} \Xi^{\alpha \mu \nu} \,, \\
\mathcal{L}_S & := S_{\alpha \mu \nu} \Sigma^{\alpha \mu \nu} \,, \\
\mathcal{L}_{QS} & := Q_{\alpha \mu \nu} \Pi^{\alpha \mu \nu} \,.
\end{split}
\end{equation}
By using this, we have 
\begin{equation}
\mathcal{L}_2 = \mathcal{L}_Q + \mathcal{L}_S + \mathcal{L}_{QS}
\end{equation}
and, consequently, the action \eqref{thy} can be rewritten as
\begin{equation}\label{thy1}
S[g,\Gamma] =\frac{1}{2 \kappa}\int d^{n}x \sqrt{-g} \Big[ R+ \mathcal{L}_Q + \mathcal{L}_S + \mathcal{L}_{QS} +\mathcal{L}_{\text{FS}} \Big] \,.
\end{equation}
Then, the variation of \eqref{thy1} with respect to the metric reads
\begin{equation}
\begin{split}
\delta_g S[g,\Gamma] & = \frac{1}{2 \kappa}\int d^{n}x \Bigg[  \delta_g \sqrt{-g} \left( R + \mathcal{L}_2 + \mathcal{L}_{\text{FS}} \right) \\
& + \sqrt{-g} \left(\delta_g R + \delta_g \mathcal{L}_Q + \delta_g \mathcal{L}_S + \delta_g \mathcal{L}_{QS} + \delta_g \mathcal{L}_{\text{FS}} \right) \Bigg] \,,
\end{split}
\end{equation}
that is
\begin{equation}
\begin{split}
\delta_g S[g,\Gamma] & = \frac{1}{2 \kappa}\int d^{n}x \Bigg[ \sqrt{-g} \left( R_{(\mu \nu)} - \frac{1}{2} g_{\mu \nu} R \right) \left(\delta g^{\mu \nu} \right) - \frac{1}{2} \sqrt{-g} g_{\mu \nu} \left( \mathcal{L}_2 + \mathcal{L}_{\text{FS}} \right) \left( \delta g^{\mu \nu} \right) \\
& + \sqrt{-g} \left( \delta_g \mathcal{L}_Q + \delta_g \mathcal{L}_S + \delta_g \mathcal{L}_{QS} + \delta_g \mathcal{L}_{\text{FS}} \right) \Bigg] \,.
\end{split}
\end{equation}
Hence, the metric field equations of the theory are
\begin{equation}\label{metricf}
\begin{split}
& R_{(\mu\nu)}-\frac{1}{2}g_{\mu\nu} \left( R + \mathcal{L}_{2} + \mathcal{L}_{\text{FS}} \right) +\frac{1}{\sqrt{-g}}\hat{\nabla}_{\alpha}\Big[ \sqrt{-g}({W^{\alpha}}_{(\mu\nu)}+{\Pi^{\alpha}}_{(\mu\nu)})\Big] + A_{(\mu\nu)} + B_{(\mu\nu)} + C_{(\mu \nu)} \\
& + 2 k_2 \Bigg \lbrace \left[ \frac{1}{2} g_{\gamma \delta} \left( \partial_\tau g^{\gamma \delta} \right) q^{\tau \beta} - \partial_\tau q^{\tau
 \beta} \right] \left( g_{(\nu | \beta} \partial^\alpha g_{\mu ) \alpha} + {\Gamma^\lambda}_{(\mu \nu)} g_{\lambda \beta} - g^{\rho \sigma} {\Gamma^\alpha}_{\rho \sigma} g_{(\mu | \alpha} g_{\nu ) \beta} \right) \\
& + \frac{1}{\sqrt{-g}} \left( \partial_{(\mu} \partial_\tau \sqrt{-g} \right) {q^\tau}_{| \nu)} - \frac{1}{2} g_{\gamma \delta} \left( \partial_\tau g^{\gamma \delta} \right) \partial_{(\mu} {q^\tau}_{| \nu)} - \frac{1}{2} g_{\gamma \delta} \left( \partial_{(\mu} g^{\gamma \delta} \right) g_{\nu ) \beta} \partial_\tau q^{\tau \beta} \\
& + \left( \partial_{(\mu} g_{\nu ) \beta} \right) \partial_\tau q^{\tau \beta} + g_{(\nu | \beta} \partial_{\mu)} \partial_\tau q^{\tau \beta} \Bigg \rbrace + 2 k_1 \hat{R}_{(\mu|\sigma} \hat{R}_{\nu)}^{\phantom{\nu)}\sigma} + 2 k_2 q_{(\mu|\sigma} {q_{\nu)}}^\sigma + 2 k_3 S_{(\mu|\sigma} {S_{\nu)}}^\sigma =0 \,,
\end{split}
\end{equation}
where we have introduce the derivative
\beq
\hat{\nabla}_{\mu}=2 S_{\mu}-\nabla_{\mu}	
\eeq
and defined
\begin{equation}
\begin{split}
{W^{\alpha}}_{(\mu\nu)} := & 2 a_{1}{Q^{\alpha}}_{\mu\nu}+2 a_{2}{Q_{(\mu\nu)}}^{\alpha}+(2 a_{3}Q^{\alpha}+a_{5}q^{\alpha})g_{\mu\nu}+(2 a_{4}q_{(\mu} + a_{5}Q_{(\mu})\delta^{\alpha}_{\nu)} \,, \\
A_{\mu\nu} := & a_{1}(Q_{\mu\alpha\beta}{Q_{\nu}}^{\alpha\beta}-2 Q_{\alpha\beta\mu}{Q^{\alpha\beta}}_{\nu})-a_{2}Q_{\alpha\beta(\mu}{Q^{\beta\alpha}}_{\nu)}+a_{3}(Q_{\mu}Q_{\nu}-2 Q^{\alpha}Q_{\alpha\mu\nu}) \\
& -a_{4}q_{\mu}q_{\nu}-a_{5}q^{\alpha}Q_{\alpha\mu\nu} \,, \\
B_{\mu\nu} := & b_{1}(2S_{\nu\alpha\beta}{S_{\mu}}^{\alpha\beta}-S_{\alpha\beta\mu}{S^{\alpha\beta}}_{\nu})-b_{2}S_{\nu\alpha\beta}{S_{\mu}}^{\beta\alpha}+b_{3}S_{\mu}S_{\nu} \,, \\
C_{\mu\nu} := & \Pi_{\mu\alpha\beta}{Q_{\nu}}^{\alpha\beta} - ( c_{1}S_{\alpha\beta\nu}{Q^{\alpha\beta}}_{\mu}+c_{2}S^{\alpha}Q_{\alpha\mu\nu}+c_{3}S^{\alpha}Q_{\mu\nu\alpha}) \\
= & c_{1}({Q_{\mu}}^{\alpha\beta}S_{\nu\alpha\beta}-S_{\alpha\beta\mu}{Q^{\alpha\beta}}_{\nu}) + c_{2}(S_{\mu}Q_{\nu}-S^{\alpha}Q_{\alpha\mu\nu}) \,,
\end{split}
\end{equation}
Taking the trace of \eqref{metricf} we obtain the following equation (note that the contribution between braces in \eqref{metricf} is traceless):
\begin{equation}\label{trace}
\Big( 1-\frac{n}{2} \Big) \left( R + \mathcal{L}_{2} \right) + \left( 2 - \frac{n}{2} \right) \mathcal{L}_{\text{FS}} - \tilde{\nabla}_{\alpha}\Big(\Pi^{\alpha}+W^{\alpha}\Big) = 0 \,,
\end{equation}
where $\tilde{\nabla}$ is the Levi-Civita covariant derivative and
\begin{equation}
\begin{split}
\Pi^{\alpha}:= & {\Pi^{\alpha}}_{\mu\nu}g^{\mu\nu}=(c_{1}+n c_{2}+c_{3})S^{\alpha} \,, \\
W^{\alpha}:= & {W^{\alpha}}_{\mu\nu}g^{\mu\nu}=(2 a_{1}+2 n a_{3}+a_{5})Q^{\alpha}+(2 a_{2} +2 a_{4}+n a_{5})q^{\alpha} \,.
\end{split}
\end{equation}
Observe that the contribution in $\mathcal{L}_{\text{FS}}$ in \eqref{trace} vanishes if one considers $n=4$ dimensions.
Plugging eq. \eqref{trace} back into \eqref{metricf}, thus eliminating the term $R + \mathcal{L}_{2}$, we are left with
\begin{equation}\label{metricf1}
\begin{split}
& R_{(\mu\nu)}-\frac{1}{n-2}g_{\mu\nu} \left[ \mathcal{L}_{\text{FS}} - \tilde{\nabla}_\alpha \left( \Pi^\alpha + W^\alpha \right) \right] +\frac{1}{\sqrt{-g}}\hat{\nabla}_{\alpha}\Big[ \sqrt{-g}({W^{\alpha}}_{(\mu\nu)}+{\Pi^{\alpha}}_{(\mu\nu)})\Big] + A_{(\mu\nu)} + B_{(\mu\nu)} \\
& + C_{(\mu \nu)} + 2 k_2 \Bigg \lbrace \left[ \frac{1}{2} g_{\gamma \delta} \left( \partial_\tau g^{\gamma \delta} \right) q^{\tau \beta} - \partial_\tau q^{\tau
 \beta} \right] \left( g_{(\nu | \beta} \partial^\alpha g_{\mu ) \alpha} + {\Gamma^\lambda}_{(\mu \nu)} g_{\lambda \beta} - g^{\rho \sigma} {\Gamma^\alpha}_{\rho \sigma} g_{(\mu | \alpha} g_{\nu ) \beta} \right) \\
& + \frac{1}{\sqrt{-g}} \left( \partial_{(\mu} \partial_\tau \sqrt{-g} \right) {q^\tau}_{| \nu)} - \frac{1}{2} g_{\gamma \delta} \left( \partial_\tau g^{\gamma \delta} \right) \partial_{(\mu} {q^\tau}_{| \nu)} - \frac{1}{2} g_{\gamma \delta} \left( \partial_{(\mu} g^{\gamma \delta} \right) g_{\nu ) \beta} \partial_\tau q^{\tau \beta} \\
& + \left( \partial_{(\mu} g_{\nu ) \beta} \right) \partial_\tau q^{\tau \beta} + g_{(\nu | \beta} \partial_{\mu)} \partial_\tau q^{\tau \beta} \Bigg \rbrace + 2 k_1 \hat{R}_{(\mu|\sigma} \hat{R}_{\nu)}^{\phantom{\nu)}\sigma} + 2 k_2 q_{(\mu|\sigma} {q_{\nu)}}^\sigma + 2 k_3 S_{(\mu|\sigma} {S_{\nu)}}^\sigma =0 \,.
\end{split}
\end{equation}
The latter relates the Ricci tensor of the Levi-Civita connection $\tilde{\Gamma}^\lambda_{\phantom{\lambda} \mu \nu}$ with the metric and the non-Riemannian quantities (i.e., the torsion and non-metricity vectors) of the theory.
Finally, one may use formulas \eqref{postrRmn}-\eqref{postrRmnS} of Appendix \ref{appa}, that is perform the post-Riemannian expansion, to further simplify the above equations, splitting Riemannian and non-Riemannian contributions. In particular, upon use of \eqref{postrR} into \eqref{trace}, the latter becomes
\begin{equation}\label{pRetrace}
\nonumber
\begin{split}
& \tilde{R} + \left( a_1 + \frac{1}{4} \right) Q_{\alpha \mu \nu} Q^{\alpha \mu \nu} + \left( a_2 - \frac{1}{2} \right) Q_{\alpha \mu \nu} Q^{\mu \nu \alpha} + \left( a_3 - \frac{1}{4} \right) Q_\mu Q^\mu + a_4 q_\mu q^\mu + \left( a_5 + \frac{1}{2} \right) Q_\mu q^\mu \\
& + \left( b_1 + 1 \right) S_{\alpha \mu \nu} S^{\alpha \mu \nu} + \left( b_2 - 2 \right) S_{\alpha \mu \nu} S^{\mu \nu \alpha} + \left( b_3 - 4 \right) S_\mu S^\mu + \left( c_1 + 2 \right) Q_{\alpha \mu \nu} S^{\alpha \mu \nu} + \left( c_2 - 2 \right) Q_\mu S^\mu \\
& + \left( c_3 + 2 \right) q_\mu S^\mu + \tilde{\nabla}_\mu \left[ \left( 2 a_2 + 2 a_4 + n a_5 + 1 \right) q^\mu + \left( 2 a_1 + 2 n a_3 + a_5 - 1 \right) Q^\mu \right] \\
& + \tilde{\nabla}_\mu \left[ \left( c_1 + n c_2 + c_3 - 4 \right) S^\mu \right] + \left( 2 - \frac{n}{2} \right) \mathcal{L}_{\text{FS}} = 0 \,.
\end{split}
\end{equation}

We conclude by observing that, considering \eqref{vanishingOmT}, the final form for the affine connection ${\Gamma^\lambda}_{\mu \nu}$ is
\begin{equation}\label{finconnection}
{\Gamma^\lambda}_{\mu\nu} = \tilde{\Gamma}^\lambda_{\phantom{\lambda}\mu\nu} + \frac12 g^{\rho\lambda}\left(Q_{\mu\nu\rho} + Q_{\nu\rho\mu}
- Q_{\rho\mu\nu}\right) - g^{\rho\lambda}\left(S_{\rho\mu\nu} +
S_{\rho\nu\mu} - S_{\mu\nu\rho}\right) \,,
\end{equation}
with
\begin{equation}\label{fintorandnonmet}
\begin{split}
Q_{\alpha\mu\nu} & = \frac{\left[(n+1)Q_\alpha - 2 q_\alpha \right]}{(n+2)(n-1)} g_{\mu \nu} + \frac{2 \left[n q_{(\mu}g_{\nu)\alpha} - Q_{(\mu}g_{\nu)\alpha}\right]}{(n+2)(n-1)} \,, \\
{S_{\lambda\mu}}^\nu & = \frac{2}{1-n} {\delta_{[\lambda}}^{\nu} S_{\mu]}\,,
\end{split}
\end{equation}
where the vectors $Q_\mu$, $q_\mu$, and $S_\mu$ obeys the Proca-like equations \eqref{procalike}. It is worth stressing out that the above connection is dynamical with the extra degrees of freedom given by the Proca fields.

All of the above describes the theory in the most general setting. Let us now focus on the particular sub-cases in which the contribution $\mathcal{L}_{\text{FS}}$ in \eqref{thy} just exhibits one of the quadratic terms in the field-strengths.

\subsection{Sub-case in which $\mathcal{L}_{\text{FS}}$ contains only the homothetic curvature}

In this case, the action \eqref{thy} reduces to
\begin{equation}\label{thyhom}
\begin{split}
S^{(1)} &=\frac{1}{2 \kappa}\int d^{n}x \sqrt{-g} \Big[ R+  
b_{1}S_{\alpha\mu\nu}S^{\alpha\mu\nu} +
b_{2}S_{\alpha\mu\nu}S^{\mu\nu\alpha} +
b_{3}S_{\mu}S^{\mu} \\
& + a_{1}Q_{\alpha\mu\nu}Q^{\alpha\mu\nu} +
a_{2}Q_{\alpha\mu\nu}Q^{\mu\nu\alpha} +
a_{3}Q_{\mu}Q^{\mu}+
a_{4}q_{\mu}q^{\mu}+
a_{5}Q_{\mu}q^{\mu} \\
& +c_{1}Q_{\alpha\mu\nu}S^{\alpha\mu\nu}+
c_{2}Q_{\mu}S^{\mu} +
c_{3}q_{\mu}S^{\mu}+ k_1 \hat{R}_{\mu \nu} \hat{R}^{\mu \nu} \Big] \\
& =\frac{1}{2 \kappa}\int d^{n}x \sqrt{-g} \Big[ R+ \mathcal{L}_{2}+ \mathcal{L}^{(1)}_{\text{FS}} \Big] \,,
\end{split}
\end{equation}
where the only non-vanishing contribution in $\mathcal{L}_{\text{FS}}$ with respect to \eqref{thy} is the one along $k_1$ ($k_1 \neq 0$), that is $\mathcal{L}^{(1)}_{\text{FS}}=k_1 \hat{R}_{\mu \nu} \hat{R}^{\mu \nu}$.

From the variation of \eqref{thyhom} with respect to the general affine connection ${\Gamma^\lambda}_{\mu \nu}$ we get
\beq
{P_{\lambda}}^{\mu\nu}+{\Psi_{\lambda}}^{\mu\nu} - 4 k_{1}\delta_{\lambda}^{\mu} D_{\alpha}\hat{R}^{\alpha\nu}=0 \,, \label{feGammak1}
\eeq
Taking the various traces of \eqref{feGammak1} we find
\begin{equation}\label{e2ap}
\begin{split}
& N_1 Q^{\nu} + N_2 q^{\nu} + N_3 S^{\nu} = 4 n k_1 D_{\alpha} \hat{R}^{\alpha \nu} \,, \\
& N_4 Q^{\mu} + N_5 q^{\mu} + N_6 S^{\mu} = 4 k_1 D_{\alpha} \hat{R}^{\alpha\mu} \,, \\
& N_7 + Q^{\lambda} + N_8 q^{\lambda} + N_9 S^{\lambda}= 4 k_1 D_{\alpha} \hat{R}^{\alpha \lambda} \,,
\end{split}
\end{equation}
where the explicit expressions of the coefficients $N_1,N_2,\ldots, N_9$ are given in Appendix \ref{appa}.
These equations can be combined in such a way to obtain
\begin{equation}\label{qSintermsofQ}
\begin{split}
q^\mu & =  A_1 Q^\mu \,, \\
S^\mu & =  A_2 Q^\mu \,,
\end{split}
\end{equation}
together with the following Proca-like equation for $Q^\mu$:
\begin{equation}\label{procaQ}
D_\alpha \hat{R}^{\alpha \mu} = A_3 Q^\mu \,,
\end{equation}
$A_3$ playing the role of the mass squared of $Q^\mu$.
The coefficients $A_1$, $A_2$, and $A_3$ are given in terms of the parameters of the theory.\footnote{Since the explicit expressions for $A_1$, $A_2$, and $A_3$ in terms of the parameters of the model are huge, we do not report them here. The key point, in fact, is that the Proca-like equations \eqref{procalike}, in the case in which only the term along $k_1$ survives, reduces to a single Proca-like equation for the non-metricity vector $Q^\mu$, without involving the other non-Riemannian vectors of the theory. The latter are indeed expressed in terms of $Q^\mu$ too (see eq. \eqref{qSintermsofQ}).}
Therefore, in this case we have just one independent non-Riemannian vector in the theory. In particular, the torsion vector $S^\mu$ and the non-metricity vector $q^\mu$ can be completely expressed in terms of the non-metricity trace $Q^\mu$, the latter obeying the Proca-like equation \eqref{procaQ}. Using all of the above into the connection field equations, we are left with \eqref{feGamma1}, which, in particular, is solved by \eqref{vanishingOmT}.

On the other hand, variation with respect to the metric field yields
\begin{equation}\label{metricfk1}
\begin{split}
& R_{(\mu\nu)}-\frac{1}{2}g_{\mu\nu} \left( R + \mathcal{L}_{2} + \mathcal{L}^{(1)}_{\text{FS}} \right) +\frac{1}{\sqrt{-g}}\hat{\nabla}_{\alpha}\Big[ \sqrt{-g}({W^{\alpha}}_{(\mu\nu)}+{\Pi^{\alpha}}_{(\mu\nu)})\Big] + A_{(\mu\nu)} + B_{(\mu\nu)} + C_{(\mu \nu)} \\
& + 2 k_1 \hat{R}_{(\mu|\sigma} \hat{R}_{\nu)}^{\phantom{\nu)}\sigma} = 0 \,,
\end{split}
\end{equation}
whose trace, in turn, gives
\begin{equation}\label{tracek1}
\Big( 1-\frac{n}{2} \Big) \left( R + \mathcal{L}_{2} \right) + \left( 2 - \frac{n}{2} \right) \mathcal{L}^{(1)}_{\text{FS}} - \tilde{\nabla}_{\alpha}\Big(\Pi^{\alpha}+W^{\alpha}\Big) = 0 \,.
\end{equation}
Finally, plugging eq. \eqref{tracek1} back into \eqref{metricfk1}, we obtain
\begin{equation}\label{metricf1k1}
\begin{split}
& R_{(\mu\nu)}-\frac{1}{n-2}g_{\mu\nu} \left[ \mathcal{L}^{(1)}_{\text{FS}} - \tilde{\nabla}_\alpha \left( \Pi^\alpha + W^\alpha \right) \right] +\frac{1}{\sqrt{-g}}\hat{\nabla}_{\alpha}\Big[ \sqrt{-g}({W^{\alpha}}_{(\mu\nu)}+{\Pi^{\alpha}}_{(\mu\nu)})\Big] + A_{(\mu\nu)} + B_{(\mu\nu)} \\
& + C_{(\mu \nu)} + 2 k_1 \hat{R}_{(\mu|\sigma} \hat{R}_{\nu)}^{\phantom{\nu)}\sigma} =0 \,.
\end{split}
\end{equation}
Note that plugging the post-Riemannian expansion \eqref{postrR} into \eqref{tracek1} the latter boils down to
\begin{equation}\label{pRetracek1}
\nonumber
\begin{split}
& \tilde{R} + \left( a_1 + \frac{1}{4} \right) Q_{\alpha \mu \nu} Q^{\alpha \mu \nu} + \left( a_2 - \frac{1}{2} \right) Q_{\alpha \mu \nu} Q^{\mu \nu \alpha} + \left( a_3 - \frac{1}{4} \right) Q_\mu Q^\mu + a_4 q_\mu q^\mu + \left( a_5 + \frac{1}{2} \right) Q_\mu q^\mu \\
& + \left( b_1 + 1 \right) S_{\alpha \mu \nu} S^{\alpha \mu \nu} + \left( b_2 - 2 \right) S_{\alpha \mu \nu} S^{\mu \nu \alpha} + \left( b_3 - 4 \right) S_\mu S^\mu + \left( c_1 + 2 \right) Q_{\alpha \mu \nu} S^{\alpha \mu \nu} + \left( c_2 - 2 \right) Q_\mu S^\mu \\
& + \left( c_3 + 2 \right) q_\mu S^\mu + \tilde{\nabla}_\mu \left[ \left( 2 a_2 + 2 a_4 + n a_5 + 1 \right) q^\mu + \left( 2 a_1 + 2 n a_3 + a_5 - 1 \right) Q^\mu \right] \\
& + \tilde{\nabla}_\mu \left[ \left( c_1 + n c_2 + c_3 - 4 \right) S^\mu \right] + \left( 2 - \frac{n}{2} \right) \mathcal{L}^{(1)}_{\text{FS}} = 0 \,.
\end{split}
\end{equation}

In the case at hand, the final form for the affine connection ${\Gamma^\lambda}_{\mu \nu}$ can be written as \eqref{finconnection}, where now
\begin{equation}
\begin{split}
Q_{\alpha\mu\nu} & = \frac{(n+1 -2 A_1)}{(n+2)(n-1)} Q_\alpha g_{\mu \nu} + \frac{(2 n A_1 - 1 )}{(n+2)(n-1)} Q_{(\mu}g_{\nu)\alpha} \,, \\
{S_{\lambda\mu}}^\nu & = \frac{2 A_2}{1-n} {\delta_{[\lambda}}^{\nu} Q_{\mu]}\,, \label{QS}
\end{split}
\end{equation}
with the non-metricity vector $Q^\mu$, that now is the only independent non-Riemannian vector, obeying the Proca-like equation \eqref{procaQ}. Interestingly enough from the last expressions we see that there exists a parameter choice for which the whole torsion vanishes (i.e., for $A_{2}=0$) but, on the other hand, there is no value for $A_{1}$ that would yield a vanishing non-metricity. Note also that for $A_{1}=\frac{1}{2 n}$ the non-metricity is restricted to be of the Weyl type.

\subsection{Sub-case in which $\mathcal{L}_{\text{FS}}$ contains only the field-strength of the $2^{nd}$ non-metricity vector $q_\mu$}

Here the action \eqref{thy} boils down to
\begin{equation}\label{thydqdq}
\begin{split}
S^{(2)} &=\frac{1}{2 \kappa}\int d^{n}x \sqrt{-g} \Big[ R+  
b_{1}S_{\alpha\mu\nu}S^{\alpha\mu\nu} +
b_{2}S_{\alpha\mu\nu}S^{\mu\nu\alpha} +
b_{3}S_{\mu}S^{\mu} \\
& + a_{1}Q_{\alpha\mu\nu}Q^{\alpha\mu\nu} +
a_{2}Q_{\alpha\mu\nu}Q^{\mu\nu\alpha} +
a_{3}Q_{\mu}Q^{\mu}+
a_{4}q_{\mu}q^{\mu}+
a_{5}Q_{\mu}q^{\mu} \\
& +c_{1}Q_{\alpha\mu\nu}S^{\alpha\mu\nu}+
c_{2}Q_{\mu}S^{\mu} +
c_{3}q_{\mu}S^{\mu}+ k_2 q_{\mu \nu} q^{\mu \nu} \Big] \\
& =\frac{1}{2 \kappa}\int d^{n}x \sqrt{-g} \Big[ R+ \mathcal{L}_{2}+ \mathcal{L}^{(2)}_{\text{FS}} \Big] \,,
\end{split}
\end{equation}
where the only non-vanishing contribution in $\mathcal{L}_{\text{FS}}$ with respect to \eqref{thy} is the one along $k_2$ ($k_2 \neq 0$), that is $\mathcal{L}^{(2)}_{\text{FS}} =k_2 q_{\mu \nu} q^{\mu \nu}$.

The connection field equation now reads
\beq
{P_{\lambda}}^{\mu\nu}+{\Psi_{\lambda}}^{\mu\nu} - 2k_{2}(g^{\mu\nu}g_{\beta\lambda}+\delta_{\beta}^{\mu}\delta_{\lambda}^{\nu})D_{\alpha}q^{\alpha\beta}=0 \,. \label{feGammak2}
\eeq
Taking the various traces of \eqref{feGammak2}, in this case we get
\begin{equation}\label{e3ap}
\begin{split}
& N_1 Q^{\nu} + N_2 q^{\nu} + N_3 S^{\nu} = 4 k_2 D_{\alpha} q^{\alpha \nu} \,, \\
& N_4 Q^{\mu} + N_5 q^{\mu} + N_6 S^{\mu} = 2 (n+1) k_2 D_{\alpha} q^{\alpha\mu} \,, \\
& N_7 Q^{\lambda} + N_8 q^{\lambda} + N_9 S^{\lambda} = 2 (n+1) k_2 D_{\alpha} q^{\alpha \lambda} \,,
\end{split}
\end{equation}
being the explicit expressions of the coefficients $N_1,N_2,\ldots, N_9$ collected in Appendix \ref{appa}.
These equations can be combined in such a way to obtain
\begin{equation}
\begin{split}
Q^\mu & = B_1 q^\mu \,, \\
S^\mu & = B_2 q^\mu \,,
\end{split}
\end{equation}
along with the Proca-like equation
\begin{equation}\label{procaq}
D_\alpha q^{\alpha \mu} = B_3   q^\mu \,,
\end{equation}
where $B_3$ plays the role of the mass squared of $q^\mu$. The coefficients $B_1$, $B_2$, and $B_3$ are given in terms of the parameters of the model. Hence, here again we have just one independent non-Riemannian vector. In particular, the torsion vector $S^\mu$ and the non-metricity vector $Q^\mu$ can be expressed in terms of $q^\mu$, where the latter fulfills the Proca-like equation \eqref{procaq}. Plugging the above results back into the connection field equations, we obtain once again \eqref{feGamma1}, solved by \eqref{vanishingOmT}.

The final form for the affine connection ${\Gamma^\lambda}_{\mu \nu}$ can be written as \eqref{finconnection}, where now we have
\begin{equation}
\begin{split}
Q_{\alpha\mu\nu} & = \frac{\left[B_1(n+1) - 2 \right]}{(n+2)(n-1)} q_\alpha g_{\mu \nu} + \frac{(2 n - B_1)}{(n+2)(n-1)} q_{(\mu}g_{\nu)\alpha} \,, \\
{S_{\lambda\mu}}^\nu & = \frac{2 B_2}{1-n} {\delta_{[\lambda}}^{\nu} q_{\mu]}\,,
\end{split}
\end{equation}
the only independent vector, $q^\mu$, fulfilling the Proca-like equation \eqref{procaq}. Similarly to the case of the previous subsection (with the inclusion of homothetic curvature) there exists a parameter configuration (namely $B_{2}=0$) for which the full torsion vanishes, while for the choice $B_{1}=2 n$ the non-metricity is restricted to be of Weyl type. Again, there is no parameter choice that makes the full non-metricity vanish.

The variation of \eqref{thydqdq} with respect to the metric yields
\begin{equation}\label{metricfk2}
\begin{split}
& R_{(\mu\nu)}-\frac{1}{2}g_{\mu\nu} \left( R + \mathcal{L}_{2} + \mathcal{L}^{(2)}_{\text{FS}} \right) +\frac{1}{\sqrt{-g}}\hat{\nabla}_{\alpha}\Big[ \sqrt{-g}({W^{\alpha}}_{(\mu\nu)}+{\Pi^{\alpha}}_{(\mu\nu)})\Big] + A_{(\mu\nu)} + B_{(\mu\nu)} + C_{(\mu \nu)} \\
& + 2 k_2 \Bigg \lbrace \left[ \frac{1}{2} g_{\gamma \delta} \left( \partial_\tau g^{\gamma \delta} \right) q^{\tau \beta} - \partial_\tau q^{\tau
 \beta} \right] \left( g_{(\nu | \beta} \partial^\alpha g_{\mu ) \alpha} + {\Gamma^\lambda}_{(\mu \nu)} g_{\lambda \beta} - g^{\rho \sigma} {\Gamma^\alpha}_{\rho \sigma} g_{(\mu | \alpha} g_{\nu ) \beta} \right) \\
& + \frac{1}{\sqrt{-g}} \left( \partial_{(\mu} \partial_\tau \sqrt{-g} \right) {q^\tau}_{| \nu)} - \frac{1}{2} g_{\gamma \delta} \left( \partial_\tau g^{\gamma \delta} \right) \partial_{(\mu} {q^\tau}_{| \nu)} - \frac{1}{2} g_{\gamma \delta} \left( \partial_{(\mu} g^{\gamma \delta} \right) g_{\nu ) \beta} \partial_\tau q^{\tau \beta} \\
& + \left( \partial_{(\mu} g_{\nu ) \beta} \right) \partial_\tau q^{\tau \beta} + g_{(\nu | \beta} \partial_{\mu)} \partial_\tau q^{\tau \beta} \Bigg \rbrace + 2 k_2 q_{(\mu|\sigma} {q_{\nu)}}^\sigma = 0 \,.
\end{split}
\end{equation}
Taking the trace of \eqref{metricfk2} we obtain
\begin{equation}\label{tracek2}
\Big( 1-\frac{n}{2} \Big) \left( R + \mathcal{L}_{2} \right) + \left( 2 - \frac{n}{2} \right) \mathcal{L}^{(2)}_{\text{FS}} - \tilde{\nabla}_{\alpha}\Big(\Pi^{\alpha}+W^{\alpha}\Big) = 0 \,.
\end{equation}
Plugging \eqref{tracek2} back into \eqref{metricfk2}, the latter becomes
\begin{equation}\label{metricf1k2}
\begin{split}
& R_{(\mu\nu)}-\frac{1}{n-2}g_{\mu\nu} \left[ \mathcal{L}^{(2)}_{\text{FS}} - \tilde{\nabla}_\alpha \left( \Pi^\alpha + W^\alpha \right) \right] +\frac{1}{\sqrt{-g}}\hat{\nabla}_{\alpha}\Big[ \sqrt{-g}({W^{\alpha}}_{(\mu\nu)}+{\Pi^{\alpha}}_{(\mu\nu)})\Big] + A_{(\mu\nu)} + B_{(\mu\nu)} \\
& + C_{(\mu \nu)} + 2 k_2 \Bigg \lbrace \left[ \frac{1}{2} g_{\gamma \delta} \left( \partial_\tau g^{\gamma \delta} \right) q^{\tau \beta} - \partial_\tau q^{\tau
 \beta} \right] \left( g_{(\nu | \beta} \partial^\alpha g_{\mu ) \alpha} + {\Gamma^\lambda}_{(\mu \nu)} g_{\lambda \beta} - g^{\rho \sigma} {\Gamma^\alpha}_{\rho \sigma} g_{(\mu | \alpha} g_{\nu ) \beta} \right) \\
& + \frac{1}{\sqrt{-g}} \left( \partial_{(\mu} \partial_\tau \sqrt{-g} \right) {q^\tau}_{| \nu)} - \frac{1}{2} g_{\gamma \delta} \left( \partial_\tau g^{\gamma \delta} \right) \partial_{(\mu} {q^\tau}_{| \nu)} - \frac{1}{2} g_{\gamma \delta} \left( \partial_{(\mu} g^{\gamma \delta} \right) g_{\nu ) \beta} \partial_\tau q^{\tau \beta} \\
& + \left( \partial_{(\mu} g_{\nu ) \beta} \right) \partial_\tau q^{\tau \beta} + g_{(\nu | \beta} \partial_{\mu)} \partial_\tau q^{\tau \beta} \Bigg \rbrace + 2 k_2 q_{(\mu|\sigma} {q_{\nu)}}^\sigma = 0 \,,
\end{split}
\end{equation}
which can be interpreted as expressing the Ricci tensor of the Levi-Civita connection in terms of the vector field $Q^{\mu}$ and its derivatives.
Finally, observe that using the post-Riemannian expansion \eqref{postrR} into \eqref{tracek2} the latter becomes
\begin{equation}\label{pRetracek2}
\nonumber
\begin{split}
& \tilde{R} + \left( a_1 + \frac{1}{4} \right) Q_{\alpha \mu \nu} Q^{\alpha \mu \nu} + \left( a_2 - \frac{1}{2} \right) Q_{\alpha \mu \nu} Q^{\mu \nu \alpha} + \left( a_3 - \frac{1}{4} \right) Q_\mu Q^\mu + a_4 q_\mu q^\mu + \left( a_5 + \frac{1}{2} \right) Q_\mu q^\mu \\
& + \left( b_1 + 1 \right) S_{\alpha \mu \nu} S^{\alpha \mu \nu} + \left( b_2 - 2 \right) S_{\alpha \mu \nu} S^{\mu \nu \alpha} + \left( b_3 - 4 \right) S_\mu S^\mu + \left( c_1 + 2 \right) Q_{\alpha \mu \nu} S^{\alpha \mu \nu} + \left( c_2 - 2 \right) Q_\mu S^\mu \\
& + \left( c_3 + 2 \right) q_\mu S^\mu + \tilde{\nabla}_\mu \left[ \left( 2 a_2 + 2 a_4 + n a_5 + 1 \right) q^\mu + \left( 2 a_1 + 2 n a_3 + a_5 - 1 \right) Q^\mu \right] \\
& + \tilde{\nabla}_\mu \left[ \left( c_1 + n c_2 + c_3 - 4 \right) S^\mu \right] + \left( 2 - \frac{n}{2} \right) \mathcal{L}^{(2)}_{\text{FS}} = 0 \,.
\end{split}
\end{equation}

\subsection{Sub-case in which $\mathcal{L}_{\text{FS}}$ contains only the field-strength of the torsion vector}

In this sub-case the theory reads
\begin{equation}\label{thydSdS}
\begin{split}
S^{(3)} &=\frac{1}{2 \kappa}\int d^{n}x \sqrt{-g} \Big[ R+  
b_{1}S_{\alpha\mu\nu}S^{\alpha\mu\nu} +
b_{2}S_{\alpha\mu\nu}S^{\mu\nu\alpha} +
b_{3}S_{\mu}S^{\mu} \\
& + a_{1}Q_{\alpha\mu\nu}Q^{\alpha\mu\nu} +
a_{2}Q_{\alpha\mu\nu}Q^{\mu\nu\alpha} +
a_{3}Q_{\mu}Q^{\mu}+
a_{4}q_{\mu}q^{\mu}+
a_{5}Q_{\mu}q^{\mu} \\
& +c_{1}Q_{\alpha\mu\nu}S^{\alpha\mu\nu}+
c_{2}Q_{\mu}S^{\mu} +
c_{3}q_{\mu}S^{\mu}+ k_3 S_{\mu \nu} S^{\mu \nu} \Big] \\
& =\frac{1}{2 \kappa}\int d^{n}x \sqrt{-g} \Big[ R+ \mathcal{L}_{2}+ \mathcal{L}^{(3)}_{\text{FS}} \Big] \,,
\end{split}
\end{equation}
where the only non-vanishing contribution in $\mathcal{L}_{\text{FS}}$ with respect to \eqref{thy} is the one along $k_3$ ($k_3 \neq 0$), that is $\mathcal{L}^{(3)}_{\text{FS}}=k_3 S_{\mu \nu} S^{\mu \nu}$.

In this case, the connection field equation is
\beq
{P_{\lambda}}^{\mu\nu}+{\Psi_{\lambda}}^{\mu\nu} - 2 k_3 D_{\alpha}S^{\alpha[\mu}\delta^{\nu]}_{\lambda} =0 \,. \label{feGammak3}
\eeq
Taking the various traces of \eqref{feGammak3}, we obtain
\begin{equation}\label{e4ap}
\begin{split}
& N_1 Q^{\nu} + N_2 q^{\nu} + N_3 S^{\nu} = (1-n) k_3 D_{\alpha} S^{\alpha \nu} \,, \\
& N_4 Q^{\mu} + N_5 q^{\mu} + N_6 S^{\mu} = (n-1) k_3 D_{\alpha} S^{\alpha\mu} \,, \\
& N_7 Q^{\lambda} + N_8 q^{\lambda} + N_9 S^{\lambda} = 0 \,,
\end{split}
\end{equation}
where the explicit expressions of the coefficients $N_1,N_2,\ldots, N_9$ are written in Appendix \ref{appa}.
These equations can be combined in such a way to obtain
\begin{equation}
\begin{split}
Q^\mu & = C_1 S^\mu \,, \\
q^\mu & = C_2 S^\mu \,,
\end{split}
\end{equation}
together with the following Proca-like equation for $S^{\mu}$:
\begin{equation}\label{procaS}
D_\alpha S^{\alpha \mu} = C_3 S^\mu \,,
\end{equation}
where $C_1$, $C_2$, and $C_3$ are given in terms of the parameters appearing in \eqref{thydSdS}. In particular, $C_3$ plays the role of the mass squared of $S^\mu$. Therefore, also in this case we have just one independent non-Riemannian vector in the theory, as the non-metricity vectors $Q^\mu$ and $q^\mu$ can be expressed in terms of the torsion vector $S^\mu$, the latter satisfying the Proca-like equation \eqref{procaS}. Again, using the equations above into the connection field equation \eqref{feGammak3} we are left with \eqref{feGamma1}, solved by \eqref{vanishingOmT}.

The metric field equations obtained from \eqref{thydSdS} read
\begin{equation}\label{metricfk3}
\begin{split}
& R_{(\mu\nu)}-\frac{1}{2}g_{\mu\nu} \left( R + \mathcal{L}_{2} + \mathcal{L}^{(3)}_{\text{FS}} \right) +\frac{1}{\sqrt{-g}}\hat{\nabla}_{\alpha}\Big[ \sqrt{-g}({W^{\alpha}}_{(\mu\nu)}+{\Pi^{\alpha}}_{(\mu\nu)})\Big] + A_{(\mu\nu)} + B_{(\mu\nu)} + C_{(\mu \nu)} \\
& + 2 k_3 S_{(\mu|\sigma} {S_{\nu)}}^\sigma =0 \,.
\end{split}
\end{equation}
Taking the trace, which yields
\begin{equation}\label{tracek3}
\Big( 1-\frac{n}{2} \Big) \left( R + \mathcal{L}_{2} \right) + \left( 2 - \frac{n}{2} \right) \mathcal{L}^{(3)}_{\text{FS}} - \tilde{\nabla}_{\alpha}\Big(\Pi^{\alpha}+W^{\alpha}\Big) = 0 \,,
\end{equation}
and substituting the latter back into \eqref{metricfk3}, we are left with
\begin{equation}\label{metricf1k3}
\begin{split}
& R_{(\mu\nu)}-\frac{1}{n-2}g_{\mu\nu} \left[ \mathcal{L}^{(3)}_{\text{FS}} - \tilde{\nabla}_\alpha \left( \Pi^\alpha + W^\alpha \right) \right] +\frac{1}{\sqrt{-g}}\hat{\nabla}_{\alpha}\Big[ \sqrt{-g}({W^{\alpha}}_{(\mu\nu)}+{\Pi^{\alpha}}_{(\mu\nu)})\Big] + A_{(\mu\nu)} + B_{(\mu\nu)} \\
& + C_{(\mu \nu)} + 2 k_3 S_{(\mu|\sigma} {S_{\nu)}}^\sigma =0 \,.
\end{split}
\end{equation}
Besides, plugging the post-Riemannian expansion \eqref{postrR} of the scalar curvature into \eqref{tracek1}, the latter yields
\begin{equation}\label{pRetracek3}
\nonumber
\begin{split}
& \tilde{R} + \left( a_1 + \frac{1}{4} \right) Q_{\alpha \mu \nu} Q^{\alpha \mu \nu} + \left( a_2 - \frac{1}{2} \right) Q_{\alpha \mu \nu} Q^{\mu \nu \alpha} + \left( a_3 - \frac{1}{4} \right) Q_\mu Q^\mu + a_4 q_\mu q^\mu + \left( a_5 + \frac{1}{2} \right) Q_\mu q^\mu \\
& + \left( b_1 + 1 \right) S_{\alpha \mu \nu} S^{\alpha \mu \nu} + \left( b_2 - 2 \right) S_{\alpha \mu \nu} S^{\mu \nu \alpha} + \left( b_3 - 4 \right) S_\mu S^\mu + \left( c_1 + 2 \right) Q_{\alpha \mu \nu} S^{\alpha \mu \nu} + \left( c_2 - 2 \right) Q_\mu S^\mu \\
& + \left( c_3 + 2 \right) q_\mu S^\mu + \tilde{\nabla}_\mu \left[ \left( 2 a_2 + 2 a_4 + n a_5 + 1 \right) q^\mu + \left( 2 a_1 + 2 n a_3 + a_5 - 1 \right) Q^\mu \right] \\
& + \tilde{\nabla}_\mu \left[ \left( c_1 + n c_2 + c_3 - 4 \right) S^\mu \right] + \left( 2 - \frac{n}{2} \right) \mathcal{L}^{(3)}_{\text{FS}} = 0 \,.
\end{split}
\end{equation}

Finally, the affine connection ${\Gamma^\lambda}_{\mu \nu}$ in the present case can be written as in \eqref{finconnection}, where now
\begin{equation}
\begin{split}
Q_{\alpha\mu\nu} & = \frac{\left[C_1(n+1) - 2 C_2 \right]}{(n+2)(n-1)} S_\alpha g_{\mu \nu} + \frac{2 ( C_2 n - C_1 )}{(n+2)(n-1)} S_{(\mu}g_{\nu)\alpha} \,, \\
{S_{\lambda\mu}}^\nu & = \frac{2}{1-n} {\delta_{[\lambda}}^{\nu} S_{\mu]} \,.
\end{split}
\end{equation}
The torsion vector $S^\mu$ is the only independent vector of the theory and obeys the Proca-like equation \eqref{procaS}.
Intriguingly, in this case there is no parameter choice that renders a vanishing torsion configuration, that is torsion is always there. On the other hand, and in contrast to the previous cases where the non-metricity field strengths were involved, here for the choice $C_{1}=0=C_{2}$ the full non-metricity vanishes. This seems to be another form of manifestation of the torsion/non-metricity duality that was reported in \cite{Iosifidis:2018zjj}. For the above parameter choice, after developing a post-Riemannian expansion and using the above results, the original action \eqref{thydSdS} is on-shell (and up to boundary terms) equivalent to 
\beq
S^{(3)}_{\text{equiv.}}=\frac{1}{2 \kappa}\int d^n x \sqrt{-g}\Big[ \tilde{R}+ k_3 S_{\mu \nu} S^{\mu \nu}+\frac{1}{2}m^{2}S_{\mu}S^{\mu} \Big] \,,
\eeq
with the mass term given by 
\beq
m^{2}=2 \frac{4+ 2b_{1}-b_{2}+(n-1)(b_{3}-4)}{(n-1)} \,,
\eeq
and where $\tilde{R}$ is the Ricci scalar of the Levi-Civita connection. The latter action represents a Vector-Tensor theory where the vector involved is $S_{\mu}$.

\section{On-shell equivalence to Vector-Tensor theories}

Let us now establish an on-shell equivalence between the Metric-Affine theory \eqref{thy} and a specific family of Vector-Tensor theories. To this end we start with the expressions \eqref{fintorandnonmet} for torsion and non-metricity. Then, computing each quadratic term and also performing a post-Riemannian expansion of the Ricci scalar, the action \eqref{thy} may be brought into the (on-shell and up to boundary terms) equivalent form 
\beq
S_{\text{equiv.}}=\frac{1}{2 \kappa} \int d^{n}x \sqrt{-g} \Big[ \tilde{R}+d_{1}A_{\mu}A^{\mu}+d_{2}A_{\mu}B^{\mu}+d_{3}B_{\mu}B^{\mu} 
+d_{4}S_{\mu}S^{\mu}+d_{5}S_{\mu}A^{\mu}+d_{6}S_{\mu}B^{\mu} + \mathcal{L}_{\text{kin}}\Big] \,, \label{equivthyds}
\eeq
where we have set $Q_{\mu}=n A_{\mu}+ 2 B_{\mu}$ and $q_{\mu}=A_{\mu}+(n+1)B_{\mu}$. The parameters $d_{i}$, $i=1,2,\ldots,6$, depend on the initial parameters of the quadratic theory and the spacetime dimensions. For completeness we report them in Appendix \ref{appa}. In terms of the torsion and non-metricity vectors the above action reads
\beq
\begin{split}
S_{\text{equiv.}} & =\frac{1}{2 \kappa} \int d^{n}x \sqrt{-g} \Big[ \tilde{R}+\frac{1}{2} m_{1}^{2}Q_{\mu}Q^{\mu}+\frac{1}{2} m_{2}^{2}q_{\mu}q^{\mu}+\frac{1}{2} m_{3}^{2}S_{\mu}S^{\mu} \\
& +\beta_{QS}Q_{\mu}S^{\mu}+\beta_{qS} S_{\mu}q^{\mu}+\beta_{Qq}Q_{\mu}q^{\mu} + \mathcal{L}_{\text{kin}}\Big] \,, \label{equivvtthy}
\end{split}
\eeq
where we have abbreviated 
\beq
\begin{split}
& m_{1}^{2}=2 \lambda_{1}^{2}\Big[ (n+1)^{2}d_{1}- (n+1)d_{2}+d_{3}\Big] \,, \\
& m_{2}^{2}=2 \lambda_{1}^{2}\Big[ 4 d_{1}- 2n d_{2}+ n^{2}d_{3} \Big] \,, \\
& m_{3}^{2}=2 d_{4} \,,
\end{split}
\eeq
with
\begin{equation}
\lambda_1 := \frac{1}{2-n(n+1)} \,,
\end{equation}
and
\beq
\begin{split}
& \beta_{QS} = \lambda_1 (d_{6}-(n+1) d_{5}) \,, \\
& \beta_{qS} = \lambda_1 (2 d_{5}-n d_{6}) \,, \\
& \beta_{Qq} = \lambda_1^2 \left[ -4(n+1)d_{1}+\Big( 2+n(n+1)\Big) d_{2}- 2 n d_{3} \right] \,,
\end{split}
\eeq
while $\mathcal{L}_{\text{kin}}$ collects all the kinetic terms along the coefficients $k_1$, $k_2$ and $k_3$.
From the above we conclude that if $m_{1}$, $m_{2}$ and $m_{3}$ are to be considered as the masses of the Proca fields, their positiveness will impose certain constraints on the parameters of the quadratic theory. In addition, the very presence of $\beta_{QS}$, $\beta_{qS}$ and $\beta_{Qq}$ indicates that the involved vector fields are, in general, interacting. However, there exists a parameter space for which all of the couplings vanish and the Proca fields become non-interacting. We shall discuss such cases in what follows. To recap this section, the quadratic MAG theory given by \eqref{thy} is on-shell equivalent to the Vector-Tensor theory \eqref{equivvtthy} consisting of three, in general, massive and interacting Proca fields existing in a Riemannian background. Let us also emphasize that one could use a field redefinition to get rid of the interaction terms. Indeed, by simply diagonalizing the matrix corresponding to the quadratic vector terms one could then define new vector fields that would be linear combinations of $S_{\mu}$, $Q_{\mu}$ and $q_{\mu}$ for which new fields no interaction term would occur. However, the price to pay in doing so is that by performing such a field redefinition we would generate unusual derivative couplings in the kinetic terms such as $\partial_{[\mu}S_{\nu]}\partial^{\mu}Q^{\nu}$, for instance. Therefore, the interactions are essential and unavoidable in the general setting where all kinetic terms are there. However, in some particular instances these interaction terms are absent, as we discuss below.

\section{On $F(R,T,Q,\mathcal{T},\mathcal{D})$ gravity plus field-strength contributions}\label{Myrzgrav}

In this section we provide an application to the case of (linear) $F(R,T,Q,\mathcal{T},\mathcal{D})$ gravity in vacuum. Before proceeding with this discussion, let us briefly sketch Metric-Affine $F(R,T,Q,\mathcal{T},\mathcal{D})$ gravity, also known as Metric-Affine Myrzakulov Gravity VIII (MAMG-VIII), and its sub-cases.

Following the idea that considering alternative geometric frameworks one can effectively gain better insights towards a deeper and more complete understanding of gravity than the one provided by GR, a rather general class of gravity theories has been developed in the literature, whose action is characterized by a generic function $F$ of non-Riemannian scalars (the scalar curvature $R$ of the general affine connection, the torsion scalar $T$, the non-metricity scalar $Q$, and, besides, the energy-momentum trace $\mathcal{T}$), which takes different form depending on the specific model. The most general of these models, which are also known as Myrzakulov gravity theories (MG-N, with $\rm{N=I,II, \ldots, VIII}$) \cite{Myrzakulov:2012ug,Harko:2021tav,Anagnostopoulos:2020lec}, is $F(R,T,Q,\mathcal{T})$ gravity, corresponding to MG-VIII. The other are sub-cases. Such theories have been generalized to the Metric-Affine framework in \cite{Iosifidis:2021kqo,Myrzakulov:2021vel}, including also a dependence on the divergence of the dilation current $\mathcal{D}$, the latter being a trace of the hypermomentum tensor \cite{Hehl:1976kt,Hehl:1976kt2,Hehl:1976kv}, in $F$.
In Table \ref{tab1} the reader can find a summary of the Metric-Affine Myrzakulov Gravity (MAMG) theories that have been developed and analyzed ($\mathcal{L}_{\text{m}}$ in Table \ref{tab1} denotes the matter action).

\begin{table}[ht]
\caption{Metric-Affine Myrzakulov Gravity theories.}\label{tab1}
\centering
\begin{tabular}{|c|c|}
\hline 
Acronym & Action \\ 
\hline 
MAMG-I & $S=\frac{1}{2k}\int d^nx \sqrt{-g}\left[F(R,T,\mathcal{D})+2k \mathcal{L}_{\text{m}}\right]$ \\ 
\hline 
MAMG-II & $S=\frac{1}{2k}\int d^nx \sqrt{-g}\left[F(R,Q,\mathcal{D})+2k \mathcal{L}_{\text{m}}\right]$\\ 
\hline 
MAMG-III & $S=\frac{1}{2k}\int d^nx \sqrt{-g}\left[F(T,Q,\mathcal{D})+2k \mathcal{L}_{\text{m}}\right]$ \\ 
\hline 
MAMG-IV & $S=\frac{1}{2k}\int d^nx \sqrt{-g}\left[F(R,T,{\cal{T}},\mathcal{D})+2k \mathcal{L}_{\text{m}}\right]$ \\ 
\hline 
MAMG-V & $S=\frac{1}{2k}\int d^nx \sqrt{-g}\left[F(R,T,Q,\mathcal{D})+2k \mathcal{L}_{\text{m}}\right]$ \\ 
\hline 
MAMG-VI & $S=\frac{1}{2k}\int d^nx \sqrt{-g}\left[F(R,Q,{\cal{T}},\mathcal{D})+2k \mathcal{L}_{\text{m}}\right]$ \\ 
\hline 
MAMG-VII & $S=\frac{1}{2k}\int d^nx \sqrt{-g}\left[F(T,Q,{\cal{T}},\mathcal{D})+2k \mathcal{L}_{\text{m}}\right]$ \\ 
\hline 
MAMG-VIII & $S=\frac{1}{2k}\int d^nx \sqrt{-g}\left[F(R,T,Q,{\cal{T}},\mathcal{D})+2k \mathcal{L}_{\text{m}}\right]$ \\ 
\hline
\end{tabular} 
\end{table}

Such theories have been analyzed mainly in four spacetime dimension and provided relevant applications in the cosmological context. For an exhaustive review of cosmological features of (MA)MG models and the way in which they offer solutions to diverse issues in the context of cosmology we refer the reader to \cite{Myrzakulov:2021vel} and references therein.
We shall now focus on $F(R,T,Q,\mathcal{T})$ gravity in some particular cases in which the function $F$ is linear in $R$, $Q$, and $T$, in vacuum, to highlight some implications of the previously discussed quadratic formulation in such model.

\subsection{Implications of the general formulation on the linear case in vacuum: $\mathrm{Model-1}$}

Let us now consider the vacuum case in which the function $F$ is linear in $R$, $Q$, and $T$, and given by the following expression:
\begin{equation}
\nonumber
F = R - Q - T - M \,,
\end{equation}
where we have defined the torsion and non-metricity scalars as
\begin{equation}\label{tornonmetscals}
\begin{split}
T & := S_{\mu \nu \alpha} S^{\mu \nu \alpha} - 2 S_{\mu \nu \alpha} S^{\alpha \mu \nu} - 4 S_\mu S^\mu \,, \\
Q & := \frac{1}{4} Q_{\alpha \mu \nu} Q^{\alpha \mu \nu} - \frac{1}{2} Q_{\alpha \mu \nu} Q^{\mu \nu \alpha} - \frac{1}{4} Q_\mu Q^\mu + \frac{1}{2} Q_\mu q^\mu \,,
\end{split}
\end{equation}
respectively. In addition, we have extended $F$ by also including the $QT$ scalar containing mixed terms and defined as
\beq
M := 2 Q_{\alpha\mu\nu}S^{\alpha\mu\nu}-2 S_{\mu}Q^{\mu}+2 S_{\mu}q^{\mu} \,.
\eeq
Hence, we are going to study the implications of the contributions along $k_1$, $k_2$, and $k_3$ into the gravitational action. Note that the combination $T+Q+M$ simply picks specific coefficients for the quadratic invariants. In particular, it corresponds to the parameter choice $a_{1}=-2 a_{2}=-a_{3}=2 a_{5}=-\frac{1}{4}$, $a_{4}=0$, $b_{1}=-1$, $b_{3}=2 b_{2}=4$ and $c_{1}=-c_{2}=c_{3}=-2$. For such an arrangement, using the post-Riemannian expansion of the Ricci scalar and the above results, our quadratic action ($\ref{thy}$) boils down to
\begin{equation}\label{thh}
S[g,\Gamma] =\frac{1}{2 \kappa}\int d^{n}x \sqrt{-g} \Big[ \tilde{R}+\mathcal{L}_{\text{FS}} \Big] \,.
\end{equation}
In other words, the quadratic terms contained in $R$ have all been canceled out by the specific combinations appearing in the sum $T+Q$. Interestingly in this case the Proca fields become massless (i.e., photon-like). Then, obviously, the action \eqref{thh} corresponds simply to GR with $3$ non-interacting ``photon'' fields associated to torsion and non-metricity. It is important to point out that the appearance of these extra fields was not imposed by hand but they rather emerged as a consequence of the generalized geometry. All three come from geometry. In conclusion, for the $F=R-T-Q-M$ case, the three Proca fields become non-interacting and massless. Let us emphasize that the reason for these last conclusions lies precisely in the fact that we have also included the mixed scalar $M$. Its absence changes the picture drastically, as we will show with the following example.

\subsection{Implications of the general formulation on the linear case in vacuum: $\mathrm{Model-2}$}

We shall now consider the linear case 
\begin{equation}
F = R - Q - T \,.
\end{equation}
In this case the post-Riemannian expansion of $R$ cancels all pure torsion and pure non-metricity scalars, but the mixed combinations remain there. In fact we have,
\beq
R-T-Q=\tilde{R}+2 Q_{\alpha\mu\nu}S^{\alpha\mu\nu}-2 S_{\mu}Q^{\mu}+2 S_{\mu}q^{\mu} \,.
\eeq
We see, therefore, that in this case the vector fields become massless but they interact with each other. That is, the simultaneous subtraction of both $T$ and $Q$ from $R$ produces massless fields.

\subsection{Implications of the general formulation on the linear case in vacuum: $\mathrm{Model-3}$}

Let us finally take the sub-case 
\begin{equation}
F = R - T \,.
\end{equation}
Now the post-Riemannian expansion of $R$ cancel all pure torsion terms leaving the torsion vector massless. Additionally, the two non-metricity vectors remain massive and all three torsion and non-metricity vectors are interacting. Similarly, for the sub-case
\begin{equation}
F = R - Q 
\end{equation}
the non-metricity vectors become massless while the torsion vector retains its mass. Interestingly, in this case the non-metricity fields do not interact with each other but only with the torsion vector.

\section{Conclusions}\label{concl}

In this paper we have extended the results presented in antecedent literature \cite{Baekler:2011jt}, considering a novel quadratic gravity action which, besides the usual Einstein-Hilbert contribution, involves all the parity even quadratic terms in torsion and non-metricity plus a Lagrangian that is quadratic in the field-strengths of the torsion and non-metricity vectors. 

In the most general case given by \eqref{thy}, in particular, we have obtained the Proca-like equations \eqref{procalike} involving all the non-Riemannian vectors of the theory, namely the torsion trace $S^\mu$ and the non-metricity vectors $Q^\mu$ and $q^\mu$. Subsequently, we have discussed the three sub-cases in which only one of the terms quadratic in the field-strengths is included into the action. When we only include the contribution quadratic in the homothetic curvature tensor, that is the field-strength of $Q^{\mu}$, the connection field equation yields a Proca-like equation for $Q^{\mu}$, while the vectors $q^\mu$ and $S^\mu$ are linearly related to the latter vector. In fact, this is in accordance with the results obtained in \cite{Obukhov:1996ka,Obukhov:1997zd}. Similarly, in the sub-case in which the only contribution quadratic in the field-strengths is given by the one involving the field-strength of the non-metricity vector $q_\mu$, we are left with a Proca-like equation for the latter, while the vectors $Q^\mu$ and $S^\mu$ are non-dynamical and proportional to $q^\mu$. Finally, the same scenario occurs if one includes into the quadratic theory just the term quadratic in the field-strength of the torsion vector $S^\mu$, and the result is a Proca-like equation for the latter. In this case, the non-metricity vectors are expressed in terms of $S^\mu$. In this latter case, there exist a consistent parameters choice for which the whole non-metricity vanishes and the action results to be on-shell equivalent (up to boundary terms) to a Vector-Tensor model in which the vector involved is $S_{\mu}$. The mass-squared contribution in the action is given in terms of the parameters of the quadratic theory. \\
Therefore, when one considers a MAG theory (in vacuum) involving the Einstein-Hilbert term for the non-Riemannian scalar curvature $R$ plus all the parity even quadratic terms in torsion and non-metricity, along with a quadratic Lagrangian contribution given in terms of just one of the field-strengths of the non-Riemannian vectors of the model, the result is an independent dynamical vector (the one whose field-strength appears in the quadratic action) fulfilling a Proca-like equation, while the other non-Riemannian vectors are non-dynamical and can be expressed in terms of the dynamical one.
We have then proved that the quadratic MAG theory given by \eqref{thy} is on-shell equivalent to the Vector-Tensor theory \eqref{equivvtthy} consisting of three, in general, massive and interacting Proca fields in a Riemannian background.

Subsequently, we have provided implications of the aforementioned formulation on the case of linear $F(R,T,Q,\mathcal{T},\mathcal{D})$ gravity in vacuum.
In particular, in the case $F=R-Q-T-M$ the quadratic terms in $R$ are canceled out by the specific combinations appearing in $T+Q$, and the theory results to be equivalent to GR with three non-interacting, massless Proca fields (that are the torsion and non-metricity vectors). On the other hand, in the case in which the mixed terms are not subtracted from $R$, that is $F=R-Q-T$, the vector fields are turn out to be massless but interacting. Finally, for $F=R-T$ the torsion vector results to be massless, while the non-metricity vectors are massive, and all three torsion and non-metricity vectors are interacting; analogously, in the case $F=R-Q$ the non-metricity vectors become massless, while the torsion vector is massive. However, in this last case, remarkably, the non-metricity vectors do not interact with each other but only with the torsion vector field.

Future developments of the present work may consist in the study of cosmological solutions of the model we have analyzed and the inclusion of matter couplings.

\section*{Acknowledgments}

D.I. acknowledges: This research is co-financed by Greece and the European Union (European Social Fund - ESF) through
the Operational Programme ``Human Resources Development, Education and Lifelong Learning'' in the context of
the project ``Reinforcement of Postdoctoral Researchers - 2nd Cycle'' (MIS-5033021), implemented by the State Scholarships Foundation (IKY). \\
L.R. would like to thank the Department of Applied Science and Technology of the Polytechnic University of Turin, and in particular Laura Andrianopoli and Francesco Raffa, for financial support. \\
This paper is supported by the Ministry of Education and Science of the Republic of Kazakhstan.

\appendix

\section{Useful formulas and coefficients in the quadratic theory}\label{appa}

For the sake of convenience, let us report in the following the explicit expressions of the coefficients appearing in eqs. \eqref{e1ap}, \eqref{e2ap}, \eqref{e3ap}, and \eqref{e4ap} of Section \ref{thetheory}:
\begin{equation}
\begin{split}
& N_1 = 4a_{1}-\frac{c_{1}}{2}+4 n a_{3}+2 a_{5}+\frac{(1-n)}{2}c_{2} \,, \\
& N_2 = 4 a_{2}+\frac{c_{1}}{2}+2 n a_{5}+4 a_{4}+\frac{(1-n)}{2}c_{3} \,, \\
& N_3 = -2b_{1}+b_{2}+2 c_{1}+2 n c_{2}+2 c_{3}+(1-n)b_{3} \,, \\
& N_4 = 2 a_{2}+\frac{c_{1}}{2}+4 a_{3}+(n+1) a_{5}+\frac{(n-1)}{2}(c_{2}-1) \,, \\
& N_5 = 4 a_{1}+2 a_{2}-\frac{c_{1}}{2}+2 a_{5}+2(n+1) a_{4}+\frac{(n-1)}{2}(c_{3}+2) \,, \\
& N_6 = 2b_{1}-b_{2}- c_{1}+2  c_{2}+(n+1) c_{3}+(n-1) b_{3}+2(2-n) \,, \\
& N_7 = 2 a_{2}+4 a_{3}+(n+1)a_{5}+\frac{(n-3)}{2} \,, \\
& N_8 = 2\Big[2 a_{1}+a_{2}+(n+1)a_{4}+a_{5}+ \frac{1}{2}\Big] \,, \\
& N_9 = 2 c_{2}-c_{1} +(n+1)c_{3}+2(n-2) \,.
\end{split}
\end{equation}
Besides, the coefficients appearing in \eqref{procalike} read
\begin{equation}\label{coeffs}
\begin{split}
& c_{11} = \frac{5 - 4 a_2 - n^2 + 4 a_1 (n+1) + 4 a_3 (n-1)(n+2)}{4 k_1 (n-1)(n+2)} \,, \\
& c_{12} = \frac{-4-8 a_1 + n (-1 + 4 a_2 + n)+ 2 a_5 (n-1)(n+2)}{4 k_1 (n-1)(n+2)} \,, \\
& c_{13} = \frac{4+c_1+c_2(n-1)-2n}{2 k_1(n-1)} \,, \\
& c_{21} = \frac{-4 - 8 a_1 + n(-1+4 a_2 + n) + 2 a_5 (n-1)(n+2)}{4 k_2 (n-1)(n+2)} = \frac{k_1}{k_2} c_{12} \,, \\
& c_{22} = \frac{1 + a_2 (n-2) + 2 a_1 n + a_4 (n-1)(n+2)}{k_2 (n-1)(n+2)} \,, \\
& c_{23} = \frac{-4 -c_1 + c_3(n-1)+2n}{2 k_2 (n-1)} \,, \\
& c_{31} = \frac{4 + c_1 + c_2(n-1)-2n}{2 k_3 (n-1)} = \frac{k_1}{k_3} c_{13} \,, \\
& c_{32} = \frac{-4 - c_1 + c_3 (n-1)+2n}{2 k_3 (n-1)} = \frac{k_2}{k_3} c_{23} \,, \\
& c_{33} = \frac{8 + 2 b_1 - b_2 + b_3 (n-1)-4n}{k_3 (n-1)} \,,
\end{split}
\end{equation}
while the $d_i$ coefficients, with $i=1,2,\ldots,6$ we have introduced in \eqref{equivthyds} are
\beq
\begin{split}
& d_{1} = - \frac{(n-1)(n-2)}{4} + (a_{1} + a_{5} + n a_{3}) + a_{2} + a_{4} \,, \\
& d_{2} = 4(a_{1}+n a_{3}) + (n+1) \Big[ 2 a_{2} + 2 a_{4} + n a_{5} + \frac{n}{2} \Big] + 2 (a_5-n) + 1 \,, \\
& d_{3} = (n+1) \Big[ 2 a_{1} + (n+1) a_{4} + 2 a_{5} \Big] +(n+3) a_{2}+ 4 a_{3} + (n-1) \,, \\
& d_{4}=\frac{4 + 2b_1 - b_2}{n-1} + b_3 - 4 \,, \\
& d_{5} = c_{1} + c_{3} + 4 + n c_2 - 2n \,, \\
& d_{6} = 2 c_2 - c_1 - 6 + (n+1)(c_3+2) \,.
\end{split}
\eeq
Finally, let us write the following post-Riemannian expansions (cf. also \cite{Iosifidis:2021tvx}):
\begin{equation}\label{postrRmn}
\begin{split}
R_{\mu \nu} & = \tilde{R}_{\mu \nu} + \Big[ - \frac{(-12-7n+2n^2+n^3)}{4[(n-1)(n+2)]^2}  Q_\lambda Q^\lambda + \frac{(-10-7n+n^3)}{2[(n-1)(n+2)]^2} Q^\lambda q_\lambda + \frac{2(n+1)}{[(n-1)(n+2)]^2} q_\lambda q^\lambda \\
& - \frac{2(n^2-5)}{n^3-3n+2} Q^\lambda S_\lambda + \frac{2(n^2-n-4)}{n^3-3n+2} q^\lambda S_\lambda - \frac{4(n-2)}{(n-1)^2} S_\lambda S^\lambda - \frac{(n+3)}{2(n-1)(n+2)} \tilde{\nabla}_\lambda Q^\lambda \\
& + \frac{(n+1)}{(n-1)(n+2)} \tilde{\nabla}_\lambda q^\lambda - \frac{2}{n-1} \tilde{\nabla}_\lambda S^\lambda \Big] g_{\mu \nu} + \frac{(n^3-7n-10)}{4[(n-1)(n+2)]^2} Q_{\mu} Q_{\nu} \\
& + \frac{4(n+1)}{[(n-1)(n+2)]^2} Q_{(\mu} q_{\nu)} - \frac{(n^2+n+2)}{[(n-1)(n+2)]^2} q_\mu q_\nu + \frac{2(n^2-n-4)}{n^3-3n+2} Q_{(\mu} S_{\nu)} \\
& + \frac{8}{n^3-3n+2} q_{(\mu} S_{\nu)} + \frac{4(n-2)}{(n-1)^2} S_{\mu} S_{\nu} + \frac{(n+1)}{2(n-1)(n+2)} \tilde{\nabla}_\mu Q_\nu - \frac{(n^2-3)}{2(n-1)(n+2)} \tilde{\nabla}_\nu Q_\mu \\
& - \frac{2}{(n-1)(n+2)} \tilde{\nabla}_{(\mu} q_{\nu)} - \frac{2(n-2)}{n-1} \tilde{\nabla}_\nu S_\mu \,,
\end{split}
\end{equation}
\begin{equation}\label{postrR}
\begin{split}
R & = \tilde{R} - \frac{(n^2-5)}{4(n-1)(n+2)} Q_\lambda Q^\lambda + \frac{(n^2-n-4)}{2(n-1)(n+2)} Q^\lambda q_\lambda + \frac{1}{(n-1)(n+2)} q_\lambda q^\lambda \\
& - \frac{2(n-2)}{n-1} Q^\lambda S_\lambda + \frac{2(n-2)}{n-1} q^\lambda S_\lambda - \frac{4(n-2)}{n-1} S^\lambda S_\lambda - \tilde{\nabla}_\lambda Q^\lambda + \tilde{\nabla}_\lambda q^\lambda - 4 \tilde{\nabla}_\lambda S^\lambda \,.
\end{split}
\end{equation}
where $\tilde{R}_{\mu \nu}$ and $\tilde{R}$ are the Ricci tensor and Ricci scalar of the Levi-Civita connection, respectively, and we recall that $\tilde{\nabla}$ is the Levi-Civita covariant derivative. Besides, we have
\begin{equation}\label{postrRmnS}
\begin{split}
R_{(\mu \nu)} & = \tilde{R}_{\mu \nu} + \Big[ - \frac{(-12-7n+2n^2+n^3)}{4[(n-1)(n+2)]^2}  Q_\lambda Q^\lambda + \frac{(-10-7n+n^3)}{2[(n-1)(n+2)]^2} Q^\lambda q_\lambda \\
& + \frac{2(n+1)}{[(n-1)(n+2)]^2} q_\lambda q^\lambda - \frac{2(n^2-5)}{n^3-3n+2} Q^\lambda S_\lambda + \frac{2(n^2-n-4)}{n^3-3n+2} q^\lambda S_\lambda - \frac{4(n-2)}{(n-1)^2} S_\lambda S^\lambda \\
& - \frac{(n+3)}{2(n-1)(n+2)} \tilde{\nabla}_\lambda Q^\lambda + \frac{(n+1)}{(n-1)(n+2)} \tilde{\nabla}_\lambda q^\lambda - \frac{2}{n-1} \tilde{\nabla}_\lambda S^\lambda \Big] g_{\mu \nu} \\
& + \frac{(n^3-7n-10)}{4[(n-1)(n+2)]^2} Q_{\mu} Q_{\nu} + \frac{4(n+1)}{[(n-1)(n+2)]^2} Q_{(\mu} q_{\nu)} - \frac{(n^2+n+2)}{[(n-1)(n+2)]^2} q_\mu q_\nu \\
& + \frac{2(n^2-n-4)}{n^3-3n+2} Q_{(\mu} S_{\nu)} + \frac{8}{n^3-3n+2} q_{(\mu} S_{\nu)} + \frac{4(n-2)}{(n-1)^2} S_{\mu} S_{\nu} + \frac{(4-n^2+n)}{2(n-1)(n+2)} \tilde{\nabla}_{(\mu} Q_{\nu)} \\
& - \frac{2}{(n-1)(n+2)} \tilde{\nabla}_{(\mu} q_{\nu)} - \frac{2(2-n)}{n-1} \tilde{\nabla}_{(\mu} S_{\nu)} \,,
\end{split}
\end{equation}
which is useful to reproduce the final results reported in the main text.

\end{document}